\documentclass[prb,twocolumn,superscriptaddress,amsmath,amssymb,showpacs]{revtex4}

\usepackage{graphicx}
\usepackage{amssymb,amsmath}
\usepackage{dcolumn}
\usepackage{bm}
\usepackage{color,soul}
\usepackage{enumerate}
\usepackage{float}

\usepackage{amsmath} 
\usepackage{amsthm} 
\usepackage{amssymb}    
\usepackage{graphicx} 
\renewcommand{\v}[1]{\ensuremath{\mathbf{#1}}} 

\newcommand{\pd}[2]{\frac{\partial #1}{\partial #2}}

\let\baraccent=\= 
\renewcommand{\=}[1]{\stackrel{#1}{=}} 

\theoremstyle{definition}

\theoremstyle{remark}

\begin{document}

\title{Vortices in spinor cold exciton condensates with spin-orbit interaction}

\author{H. Sigurdsson}
\affiliation{Division of Physics and Applied Physics, Nanyang
Technological University 637371, Singapore} \affiliation{Science
Institute, University of Iceland, Dunhagi-3, IS-107, Reykjavik,
Iceland}

\author{T. C. H. Liew}
\affiliation{Division of Physics and Applied Physics, Nanyang Technological University 637371, Singapore}

\author{O. Kyriienko}
\affiliation{Division of Physics and Applied Physics, Nanyang Technological University 637371, Singapore}
\affiliation{Science Institute, University of Iceland, Dunhagi-3, IS-107, Reykjavik, Iceland}

\author{I. A. Shelykh}
\affiliation{Division of Physics and Applied Physics, Nanyang Technological University 637371, Singapore}
\affiliation{Science Institute, University of Iceland, Dunhagi-3, IS-107, Reykjavik, Iceland}

\date{\today}

\begin{abstract}
We study theoretically the ground states of topological defects in a spinor four-component condensate of cold indirect excitons. We analyze possible ground state solutions for different configurations of vortices and half-vortices. We show that if only Rashba or Dreselhaus spin-orbit interaction (SOI) for electrons is present the stable states of topological defects can represent a cylindrically symmetric half-vortex or half vortex-antivortex pairs, or a non-trivial pattern with warped vortices. In the presence of both of Rashba and Dresselhaus SOI the ground state of a condensate represents a stripe phase and vortex type solutions become unstable.
\end{abstract}

\pacs{71.35.Lk,03.75.Mn,71.70.Ej}
\maketitle

\section{Introduction}
An existence of topological phases and excitations can be seen as a
manifestation of unique and universal laws of physics. Being widely
studied in various systems, a remarkable understanding of
topological defects was attained for Bose-Einstein condensates of
ultracold atoms,\cite{Anderson,Davis} where quantization of angular
momentum was experimentally observed.\cite{Matthews} The resulting
quasiparticles --- quantum vortices --- consist of a vortex core,
where the condensate density reaches its minimum and phase becomes
singular, and a circulating superfluid flow around, with phase
winding being an integer number of $2\pi$.\cite{ReviewBEC} Other
examples of topological defects include domain walls,
solitons,\cite{Kivshar} warped vortices,\cite{RuboComment}
skyrmions,\cite{Stoof} and fractional vortices which can appear in
multicomponent\cite{Babaev} or spinor condensate
systems.\cite{Vollhardt}

The usual scheme for generation of vortices in atomic physics is
based on the effective Lorentz force appearing due to rotation of
the condensate.\cite{Dalibard2000,Ketterle2001} Recently an
alternative approach with an optically-induced artificial gauge
field generation was realized.\cite{Lin2009} The next important step
forward in manipulation of atomic condensates was performed with
implementation of an artificial spin-orbit coupling between several
spin components.\cite{Lin2011} Followed by numerous theoretical
proposals,\cite{Galitski,Wu,Ramachandhran} this system was shown to
be an excellent playground for studying diverse spin-related
topological phases and excitations,\cite{Galitski2013} including
single plane wave and striped phases,\cite{Wang2010}
hexagonally-symmetric phase,\cite{Sinha2011} square vortex
lattice,\cite{Ruokokoski2012} skyrmion lattice,\cite{HuiHu2012} and
even a quasicrystalline phase for cold dipolar
bosons.\cite{Gopalakrishnan2013}

A major drawback in the study of cold atom systems is the ultralow
temperature ($< 1$ nK) required for condensation of atoms in
magnetic traps. However, solid-state physics offers a large variety
of systems, where bosonic quasiparticles with small effective mass
can condense at comparably high temperatures. They include Quantum
Hall bilayers,\cite{Eisenstein} magnons,\cite{Democritov}, indirect
excitons,\cite{Butov2001,SnokeScience,High2012} and cavity
exciton-polaritons.\cite{KasprzakNature,Balili,BaumbergPRL2008,SvenNature}
Moreover, the latter system possesses a spinor structure being
formed by two polariton spin components with $\pm 1$ spin
projection.\cite{ShelykhSpinopt} Complementary to the full quantum
vortices in the polariton
fluids,\cite{Lagoudakis2008,Liew2008,Lagoudakis2011} this allows one
to study half-integer quantum
vortices\cite{Rubo2007,Lagoudakis2009,Manni2012} and their warped
analogs.\cite{Rubo2007,Flayac2010,RuboComment} An intriguing spin
dynamics there is caused by an analog of spin-orbit interaction
(SOI) given by momentum-dependent TE-TM
splitting.\cite{Shelykh2006,Liew2007,Manni2011}

Even higher spin degeneracy can be achieved for the system of
indirect excitons --- bound pairs of electrons and holes which are
spatially separated in two parallel quantum wells [Fig.
\ref{Fig1}].\cite{Lozovik,Timofeev,Butov2007} Due to the small
overlap between the wavefunctions of electrons and holes these
quasiparticles possess very large radiative lifetime (up to
microseconds), which allows them to thermalize and consequently form
a macroscopically coherent state with properties similar to a
Bose-Einstein condensate.\cite{High2012} Another important feature
of indirect exciton gases is formation of a so-called
macroscopically ordered state manifesting itself as a fragmented
exciton ring.\cite{Butov2002,Snoke2002,Dubin2012}

Accounting for four possible $\pm 1, \pm 2$ cold indirect exciton
spin projections, an ambiguous choice of condensate ground state is
possible.\cite{Rubo2011,Kyriienko2012} This results in non-trivial
condensate topology and the possibility for generation of various
topological defects.\cite{Matuszewski2012} Moreover, complex spin
textures around fragmented beads of cold exciton condensates were
observed.\cite{High2013} They were explained with an influence of
SOI of various types, which affects the center-of-mass exciton
motion.\cite{Kyriienko2012,Vishnevsky2013,Kavokin2013} This assures
that physics similar to atomic spin-orbit coupled condensates,
including artificial magnetic field
generation,\cite{Arnardottir2012} can be studied with cold indirect
excitons.

In this paper we study the ground states of various topological
defects in an indirect exciton condensate. We show that the presence
of the SOI leads to drastic changes in the ground state of the
topological defects in the indirect exciton condensate. Using the
imaginary-time Gross-Pitaevskii equations for the spinor macroscopic
wave function, we find that in the presence of only one type of SOI
half-vortex solutions are possible, while for both Rashba and
Dresselhaus SOI the  only possible stable solution is a striped
state with zero vorticity. We study the numerical solutions of the
equations and derive analytical estimates for the boundaries, which
define topological charge stabilities. The results are consistent
with recent experimental observations of spin textures in a diluted
coherent gas of cold indirect excitons.

\section{The model}

\begin{figure}
\includegraphics[width=1.0\linewidth]{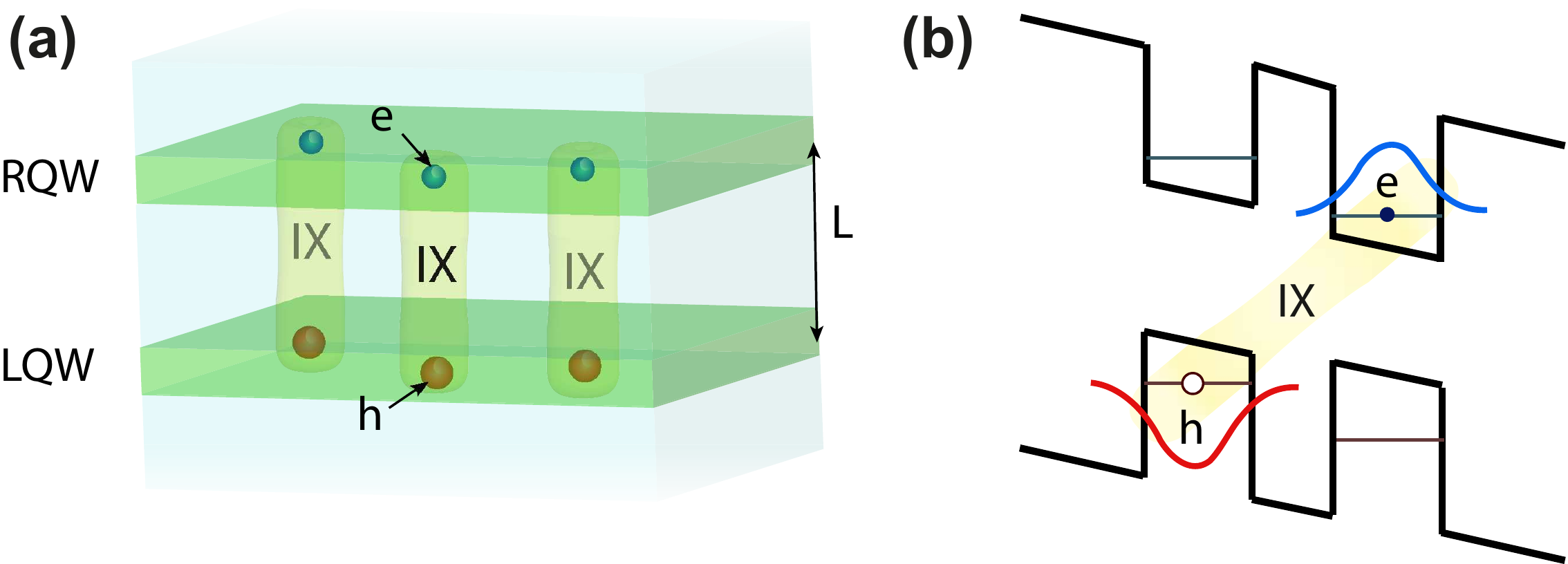}
\caption{(Color online) Sketch of the system. (a) A heterostructure
with biased coupled quantum wells, where an electron from the right
quantum well (RQW) is coupled with a heavy hole in the left quantum
well (LQW), forming an indirect exciton (IX). (b) The energy
structure of an electron-hole bilayer showing spatial separation of
electron (e) and hole (h) wave functions.} \label{Fig1}
\end{figure}

An indirect exciton is a composite boson consisting of a spatially
separated electron and hole [Fig. \ref{Fig1}]. Its spin is defined
by electron and heavy hole spin projections on the structure growth
axis, being $\pm 1/2$ and $\pm 3/2$, respectively. The resulting
four combinations correspond to possible exciton spin projections,
$S_{z}=\pm 1, \pm 2$. The states with $S_{z}=\pm 1$ are called the
bright excitons, since they can be optically excited by an external
pump. In contrast, the states with $S_{z}=\pm 2$ spin are optically
inactive due to angular momentum conservation selection rules and
are typically referred to as dark excitons. However, they can appear
due to exchange interaction between bright states or as a result of
spin-orbit interaction. In the case of direct excitons the bright
and dark states are typically split by short range electron-hole
exchange, with dark states lying at lower energies.\cite{Maialle}
This can possibly lead to the dark or gray condensation in the
corresponding systems, which prevents direct observation of
macroscopic coherence in the photoluminescence
measurements.\cite{Combescot2007,Combescot2012} Moreover, the
effects of spin-orbit interactions where shown to interplay with a
bright-dark splitting, leading to unconventional pairing effects in
the dense BCS-like direct exciton condensates.\cite{Can2009} In the
case of indirect excitons the small overlap between electron and
hole wave functions leads to approximately equal energies of all
four indirect exciton states. The dark states still play an
important role and cannot be excluded from the
consideration.\cite{Combescot, CombescotPRB}

To describe a coherent state of indirect excitons, we can use the
mean-field treatment similar to Refs.
[\onlinecite{Kyriienko2012,Matuszewski2012}], where the
Gross-Pitaevskii equation for the four-component wave function
$\Psi=\left(\Psi_{+2},\Psi_{+1},\Psi_{-1},\Psi_{-2}\right)$ was
introduced. In the general form it can be derived varying the
Hamiltonian density over the macroscopic wave function, $i \hbar
d\Psi/dt = \partial \mathcal{H}/\partial \Psi^*$. The Hamilton
density can be written as a sum of a linear single particle and
nonlinear interaction parts, $\mathcal{H} = \mathcal{H}_0 +
\mathcal{H}_{\text{int}}$.

The single particle part of the Hamiltonian density is composed of
the kinetic energy and SOI.  The latter appears as a consequence of
spin-orbit interaction acting on the electron or hole spin. In the
following we will account only for the part of SOI affecting the
spin of electron. It consists of two terms. The Dresselhaus term
arises from bulk inversion asymmetry and for a [001] quantum well is
described by the Hamiltonian $H_{D}=\beta(\sigma_x k_x - \sigma_y
k_y)$, where $k_{x,y}$ are Cartesian components of the electron wave
vector, $\sigma_{x,y}$ are Pauli matrices, and $\beta$ denotes the
strength of the Dresselhaus interaction. The Rashba term appears due
to structure inversion asymmetry and is described by the Hamiltonian
$H_{R}= \alpha(\sigma_x k_y - \sigma_y k_x)$, with $\alpha$ being
the strength of the Rashba interaction.

The single particle term in the Hamiltonian density thus reads
\cite{Kyriienko2012}:
\begin{equation}
\mathcal{H}_0 = \Psi^\dagger \hat{\mathbf{T}} \Psi,
\end{equation}
where
\begin{equation}
\label{eq.hammatrix}
\hat{\mathbf{T}} = \begin{pmatrix} \hat{\mathbf{T}}_0 & \emptyset \\ \emptyset & \hat{\mathbf{T}}_0 \end{pmatrix},
\end{equation}
with $\emptyset$ being a null matrix and
\begin{equation}
\label{eq.hammatrix0}
\hat{\mathbf{T}}_0 = \begin{pmatrix} \hbar^2 K^2 / 2m_X & S_K \\  S_K^* & \hbar^2 K^2 / 2m_X \end{pmatrix}.
\end{equation}
Here
\begin{equation} \label{eq.ksoi}
S_K = \chi [ \beta (K_x + i K_y)  +  \alpha (K_y + i K_x) ],
\end{equation}
where $\chi = m_e/m_X$ is the electron-to-exciton mass ratio and $K
= -i \nabla$ denotes the center of mass wave vector of the indirect
exciton. Note that in the described Hamiltonian we neglect the bright-dark splitting of the indirect exciton states. This however can be straightforwardly introduced for the systems, where such a splitting was observed.\cite{Combescot2012,Kavokin2013}

The nonlinear part of the Hamiltonian density $H_{\text{int}}$
describes interactions between indirect excitons. Since excitons are
composite bosons, there are four possible types of interactions
corresponding to the exchange of electrons ($V_e$), exchange of
holes ($V_h$), simultaneous exchange of electron and hole (or
exciton exchange, $V_X$), and direct Coulomb repulsion ($V_{dir}$).
Introducing the interaction constants $V_0 = V_e + V_h +
V_{\text{dir}} + V_X$ and $W = V_e + V_h$, the interaction part of
the Hamiltonian density becomes
\begin{align} \label{eq.hamdens} \notag
\mathcal{H}_{\text{int}} & =\frac{V_0}{2}\left(|\Psi_{+2}|^2+|\Psi_{+1}|^2+|\Psi_{-1}|^2+|\Psi_{-2}|^2\right)^2 \\
&+W\left(\Psi^*_{+1}\Psi^*_{-1}\Psi_{+2}\Psi_{-2}+\Psi^*_{+2}\Psi^*_{-2}\Psi_{+1}\Psi_{-1}\right) \\ \notag
&-W\left(|\Psi_{+2}|^2|\Psi_{-2}|^2+|\Psi_{+1}|^2|\Psi_{-1}|^2\right).
\end{align}

We mainly focus on the weakly depleted Bose-Einstein condensates of
indirect excitons where the biggest interaction contribution comes
from vanishing transferred momentum $q$, thus working in the long
wavelength limit $(q \to 0)$, where $V_{\text{dir}} = V_X$ and $V_e
= V_h$ (s-wave approximation). The interaction parameters can be
further estimated using a narrow QW
approximation.\cite{Kyriienko2012}

As a specific system corresponding to our model we consider the
indirect exciton system studied in Ref. [\onlinecite{High2012}],
where macroscopic coherence of indirect exciton gas was reported.
The studied sample is high quality double quantum well structure
with $8$ nm GaAs QWs and a $4$ nm Al$_{0.33}$Ga$_{0.67}$As barrier.
The observation of nontrivial spin structures in the same sample
presumes an importance of spin-orbit interaction in the described
system.\cite{High2013}

The dynamics of the system is described by a set of four coupled
nonlinear equations of Gross-Pitaevskii type:
\begin{widetext}
 \begin{equation} \label{eq.gp1} i \hbar \pd{\Psi_{+2}}{t} = \hat{E}\Psi_{+2} + \hat{S}_{R} \Psi_{+1} + V_0 \Psi_{+2} |\Psi_{+2}|^2 + (V_0 - W) \Psi_{+2} |\Psi_{-2}|^2 + V_0 \Psi_{+2}( |\Psi_{-1}|^2 + |\Psi_{+1}|^2) + W \Psi_{-2}^* \Psi_{+1} \Psi_{-1}, \end{equation}
 \begin{equation} \label{eq.gp2}
i \hbar \pd{\Psi_{+1}}{t} = \hat{E}\Psi_{+1} - \hat{S}_{R}^* \Psi_{+2} + V_0 \Psi_{+1} |\Psi_{+1}|^2 + (V_0 - W) \Psi_{+1} |\Psi_{-1}|^2 + V_0\Psi_{+1}( |\Psi_{-2}|^2 + |\Psi_{+2}|^2) + W \Psi_{-1}^* \Psi_{+2} \Psi_{-2},
\end{equation}
 \begin{equation} \label{eq.gp3} i \hbar \pd{\Psi_{-1}}{t} = \hat{E}\Psi_{-1} + \hat{S}_{R} \Psi_{-2} + V_0 \Psi_{-1} |\Psi_{-1}|^2 + (V_0 - W) \Psi_{-1} |\Psi_{+1}|^2 + V_0 \Psi_{-1} ( |\Psi_{+2}|^2 + |\Psi_{-2}|^2) + W \Psi_{+1}^* \Psi_{+2} \Psi_{-2}, \end{equation}
 \begin{equation} \label{eq.gp4} i \hbar \pd{\Psi_{-2}}{t} = \hat{E}\Psi_{-2} - \hat{S}_{R}^* \Psi_{-1} + V_0 \Psi_{-2} |\Psi_{-2}|^2 + (V_0 - W) \Psi_{-2} |\Psi_{+2}|^2 + V_0 \Psi_{-2}( |\Psi_{+1}|^2 + |\Psi_{-1}|^2) + W \Psi_{+2}^* \Psi_{+1} \Psi_{-1}. \end{equation}
\end{widetext}
Here $\hat{E} = -\hbar^2 \nabla^2 /2m_X$ is the exciton kinetic energy operator and
\begin{equation}
\hat{S}_R = \chi \left[ \beta (\hat{\partial}_y - i \hat{\partial}_x) + \alpha (\hat{\partial}_x - i \hat{\partial}_y) \right]
\end{equation}
is the SOI operator accounting for both Rashba $(\alpha)$ and
Dresselhaus $(\beta)$ contributions.

\section{Numerical method}
We use the imaginary time method to find the state corresponding to
the local minima of the Hamiltonian of the interacting exciton
system described by Eqs. (\ref{eq.gp1})-(\ref{eq.gp4}). Fourier
spectral methods are used in space and a variable order
Adams-Bashforth-Moulton method in time to achieve accurate discrete
gradient flow towards a possible low energy state. Note, that the
energy profile can have multiple minima, and the one that is reached
in numerical procedure strongly depends on the initial conditions.
In particular, one can suppose that if a stable vortex is present in
the system the corresponding solution will be found if the initial
distribution contains non-zero vorticity, while the ground state
with zero vorticity (homogeneous or striped) will be found if one
does not have an initial angular momentum. If the system does not
possess any stable solutions in the form of vortices, state with no
angular momentum will be recovered independently of the initial
condition.

We introduce a weak harmonic 2D-trapping potential $V_{\text{trap}}$ in the Hamiltonian to keep the condensate localized
within the system. The trap profile is given by $V_{\text{trap}} =
u_{0} \v{r}_{\perp}^2$, where $u_{0}=m_X \omega^2/2$ represents the
trap strength.

A choice of initial conditions is not always trivial when dealing
with a nonlinear set of equations controlled by many parameters. In
our case the typical initial condition corresponds to the vortex
solution:
\begin{equation} \label{eq.psiin}
\Psi_\sigma(\v{r})= R^{(0)}(r) \frac{r/\chi}{\sqrt{ r^2/\chi^2 + 1}} e^{i (m_\sigma \theta + k_{\sigma} \pi)}.
\end{equation}
Here $R^{(0)}(r)$ is a Gaussian function corresponding to the trapped exciton gas, $\sigma$ is the spin index and $\chi$ is the healing length of the vortex in a one component BEC given by $\chi = \hbar/\sqrt{2 m_X
V_0 n}$, \cite{Fetter Svidzinsky} where $V_0$ is the nonlinear
interaction parameter defined before and $n$ is the 2D density of
the exciton gas. The effective mass of the exciton is taken to be
$m_X = 0.21m_e$, where $m_e$ is the free electron
mass.\cite{Butov2002} We assume that the healing length of a vortex in a four component BEC is comparable with one component BEC case.

We stress that the initial condition is used here only to set
different topologies in the system. The final result of imaginary
time propagation obtains the minimum energy state for a given
topology (if such a state exists), that is, the ground state of a given topological defect characterized by winding numbers $m_\sigma$. We checked that such
solutions are unchanged for different topologically invariant
spatial profiles of the initial conditions; changing the specific
shape of the radial wave function does not change the final result.
One could start the calculation with just uniform density subject to
some circulation and the density dip of the vortex appears in the
ground state results.

Note that the relative phases between the components in the initial condition (set by $k_\sigma$) can affect the solution. Where this is so, we minimize over different values of $k_\sigma$ to find the minimum energy state. Finally, we confirm that our results are stationary states by propagating them in real time numerically.

\section{Vortices, half vortices, and half vortex-antivortex pairs}

Let us first consider the cylindrically symmetric stationary wave
function of the Gross-Pitaevskii equation as a possible minimal
energy state for a rotating BEC around the $z$-axis,\cite{Fetter
Svidzinsky}
\begin{equation} \label{eq.stat}
\Psi_\sigma(r,\theta,t) = R_{\sigma}(r) e^{i (m_{\sigma} \theta + k_\sigma \pi)} e^{-i \mu t},
\end{equation}
where $\mu$ is the chemical potential of the condensate. The
circulation of the tangential velocity over a closed contour for
quantum vortices is quantized in units of $2\pi \hbar/m_X$
controlled by the winding number $m_{\sigma}$, also known as
\emph{vorticity} or \emph{topological charge}. Recent works on
spinor exciton condensates have concluded that one of the simplest
vortex solutions is of opposite vorticity in the $\Psi_{\pm 1}$
components (half vortex-antivortex pair) and zero vorticity in the
dark components (or vice versa).\cite{Matuszewski2012, Flayac2010}
This will later be shown to be indeed a possible low energy solution
amongst other interesting vortex solutions for different
$m_{\sigma}$ and $k_{\sigma}$.

The radial part is taken to be purely real and is related to the total density of the condensate as
\begin{equation}
|R_{+1}|^2  + |R_{-1}|^2 + |R_{+2}|^2 + |R_{-2}|^2 = n,
\end{equation}
where
\begin{equation} \label{eq.totpar}
\int \sum_{\sigma} |R_{\sigma}|^2 \ d^2 r = N
\end{equation}
is the total number of excitons in the system. In this paper we use the exciton density in the harmonic
trap being $n \propto 10^8$ cm$^{-2}$. The lateral size of the system of $20~\mu$m was chosen corresponding to localized bright spots observed in past experiments on exciton condensates.\cite{High2013} The total number of particles was estimated as $N \approx 100$.

The phase difference $k_\sigma \pi$ becomes essential in whether the
vortex solution is present in the condensate or not. Adding $\pi$
phase difference switches the sign of the wave function and thus
switches the sign of the second line term in the nonlinear part of
the Hamiltonian density [Eq. (\ref{eq.hamdens})] corresponding to
bright to dark exciton conversion. Moreover, Eq. (\ref{eq.stat})
reveals that for the solution to be cylindrically symmetric in the
spinor exciton condensate the winding numbers need to satisfy the
following bound:
\begin{equation} \label{eq.bound1}
m_{+1} + m_{-1} = m_{+2} + m_{-2}.
\end{equation}
Let us rewrite Eqs. (\ref{eq.gp1})-(\ref{eq.gp4}), in the limit that the SOI strength is zero:
 \begin{equation} \label{eq.gp5}
 i \hbar \pd{\Psi_{+2}}{t} = \hat{E}\Psi_{+2} + V_0 n \Psi_{+2} + W \Psi_{-2}^* \Psi_{\Delta}^2,
\end{equation}
 \begin{equation} \label{eq.gp6}
i \hbar \pd{\Psi_{+1}}{t} = \hat{E}\Psi_{+1} + V_0 n \Psi_{+1} - W \Psi_{-1}^* \Psi_{\Delta}^2,
\end{equation}
\begin{equation} \label{eq.gp7}
 i \hbar \pd{\Psi_{-1}}{t} = \hat{E}\Psi_{-1} + V_0  n \Psi_{-1}- W  \Psi_{+1}^* \Psi_{\Delta}^2,
\end{equation}
 \begin{equation} \label{eq.gp8}
 i \hbar \pd{\Psi_{-2}}{t} = \hat{E}\Psi_{-2} + V_0 n \Psi_{-2} + W \Psi_{+2}^* \Psi_{\Delta}^2,
\end{equation}
where we used definition $\Psi_{\Delta}^2 \equiv
\Psi_{+1}\Psi_{-1}-\Psi_{+2}\Psi_{-2}$. Eqs.
(\ref{eq.gp5})-(\ref{eq.gp8}) show that the only difference between
the equations describing bright and dark excitons is the sign of the
$W$ term describing bright to dark exciton conversion. This symmetry
between bright and dark components means that if topological defects
exist for the bright excitons then the same defects can exist for
the dark excitons.
\begin{figure}[h!]
\centering

  \begin{minipage}[b]{0.48\linewidth}
    \centering
    \includegraphics[width=\linewidth]{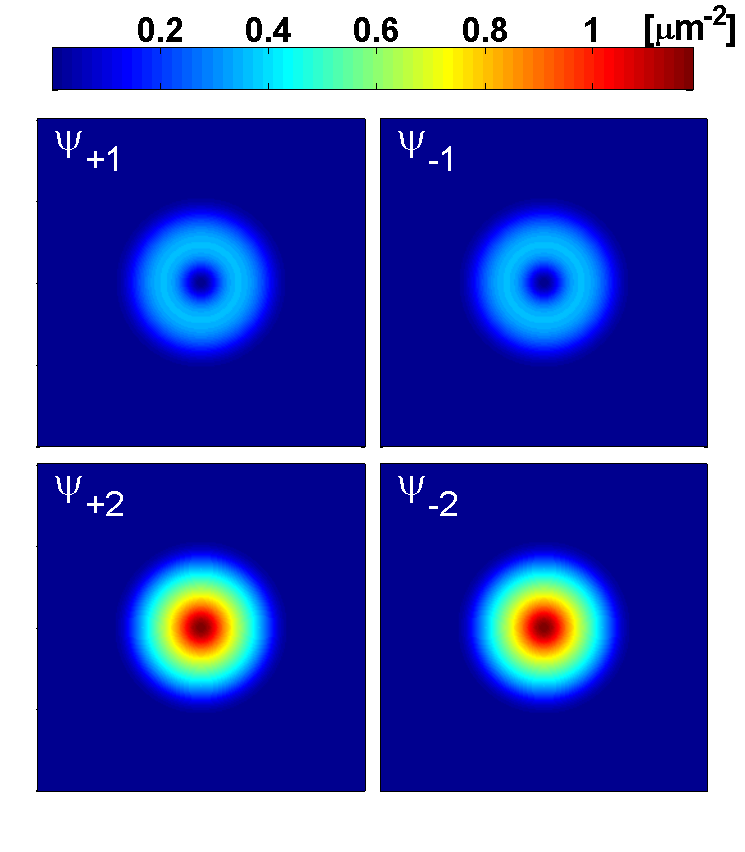}
  \end{minipage}
  \hspace{0cm}
  \begin{minipage}[b]{0.48\linewidth}
    \centering
    \includegraphics[width=\linewidth]{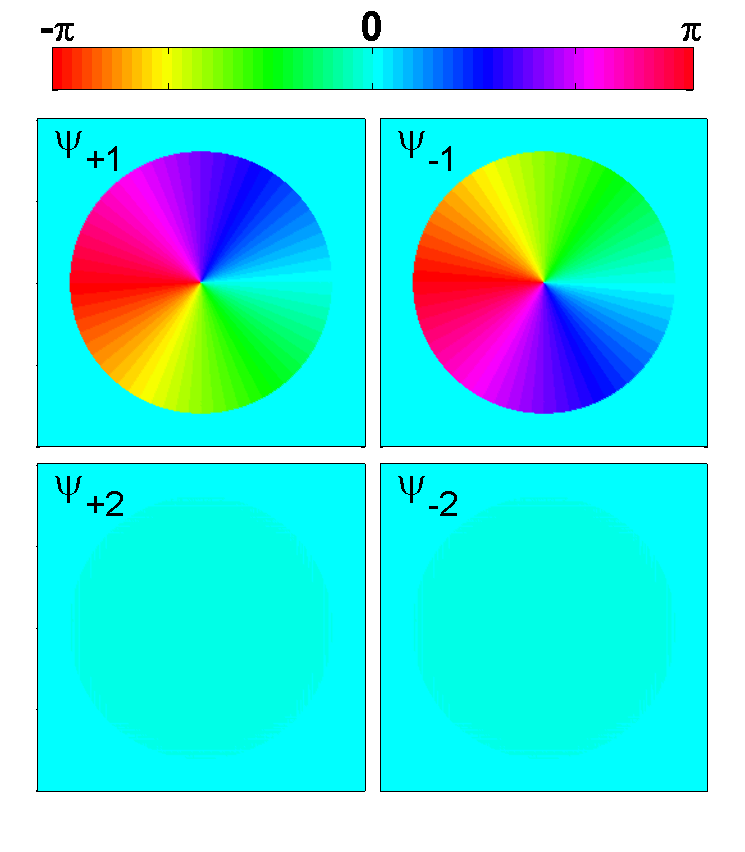}
  \end{minipage}

\centering
  \begin{minipage}[b]{0.48\linewidth}
    \centering
    \includegraphics[width=\linewidth]{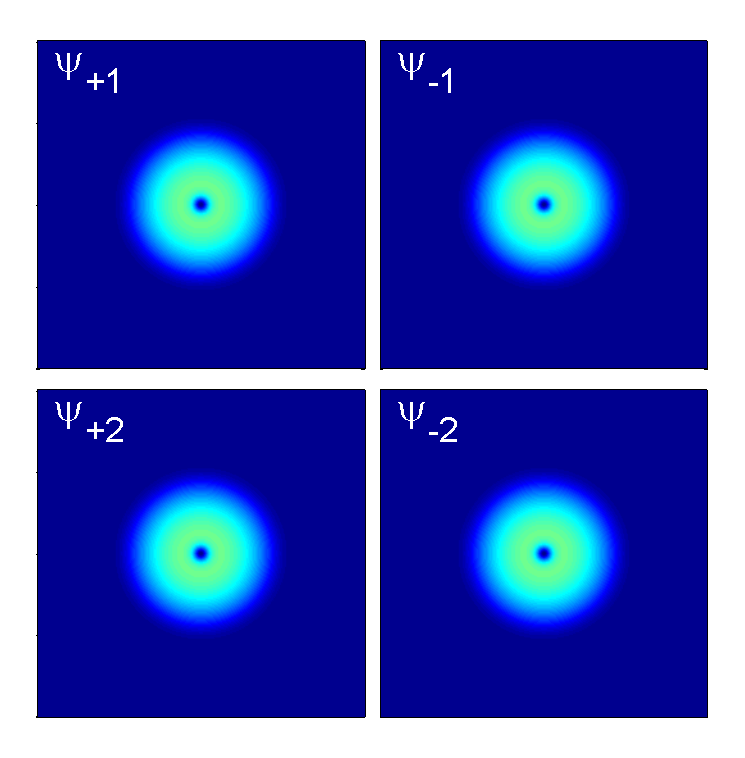}
  \end{minipage}
  \hspace{0cm}
  \begin{minipage}[b]{0.48\linewidth}
    \centering
    \includegraphics[width=\linewidth]{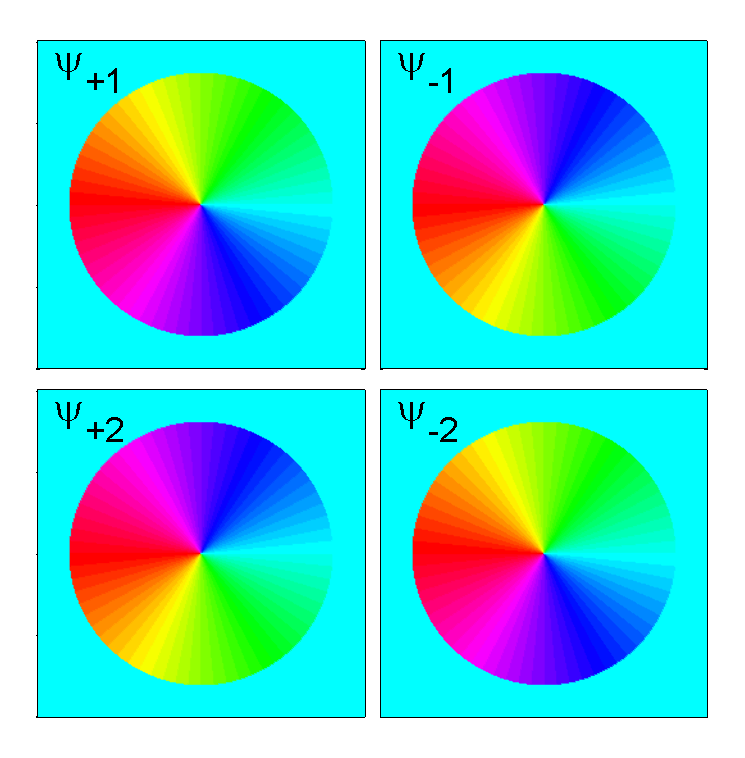}
  \end{minipage}

\centering
  \begin{minipage}[b]{0.48\linewidth}
    \centering
    \includegraphics[width=\linewidth]{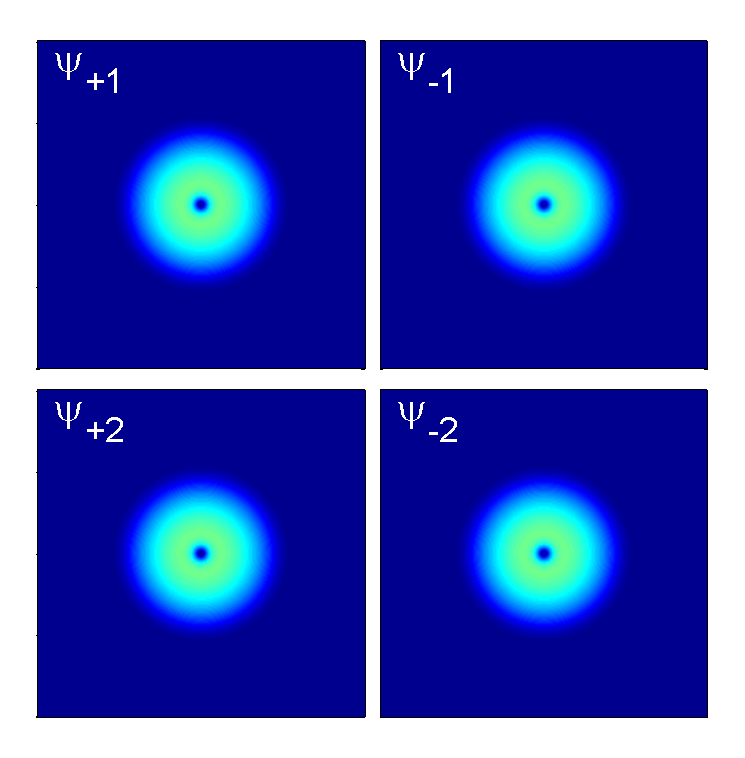}
  \end{minipage}
  \hspace{0cm}
  \begin{minipage}[b]{0.48\linewidth}
    \centering
    \includegraphics[width=\linewidth]{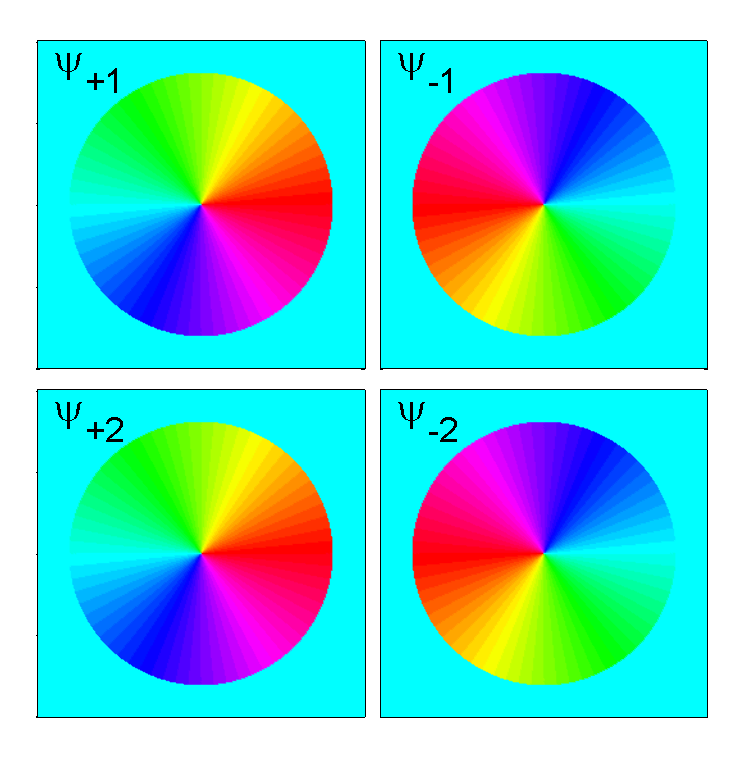}
  \end{minipage}

\centering
  \begin{minipage}[b]{0.48\linewidth}
    \centering
    \includegraphics[width=\linewidth]{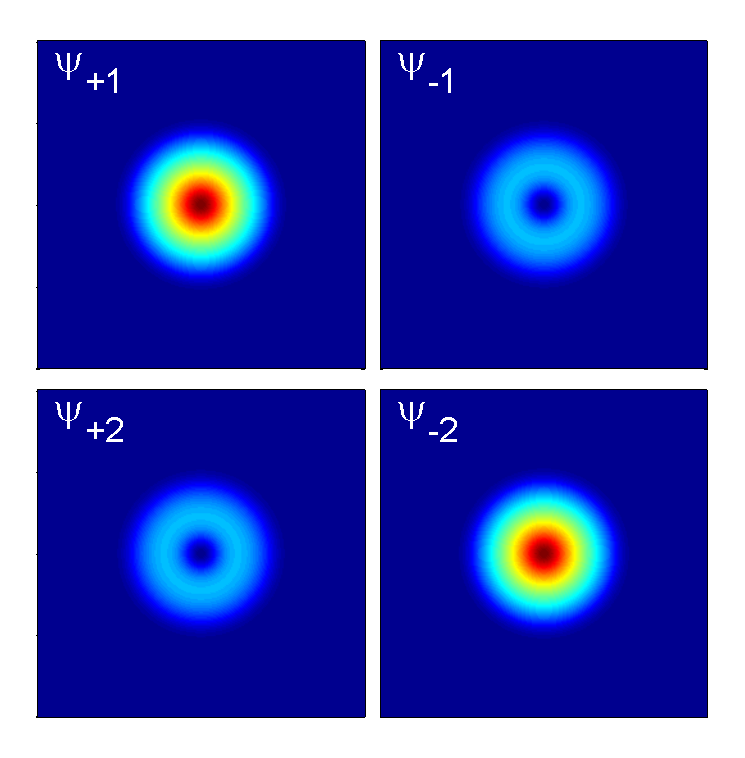}
  \end{minipage}
  \hspace{0cm}
  \begin{minipage}[b]{0.48\linewidth}
    \centering
    \includegraphics[width=\linewidth]{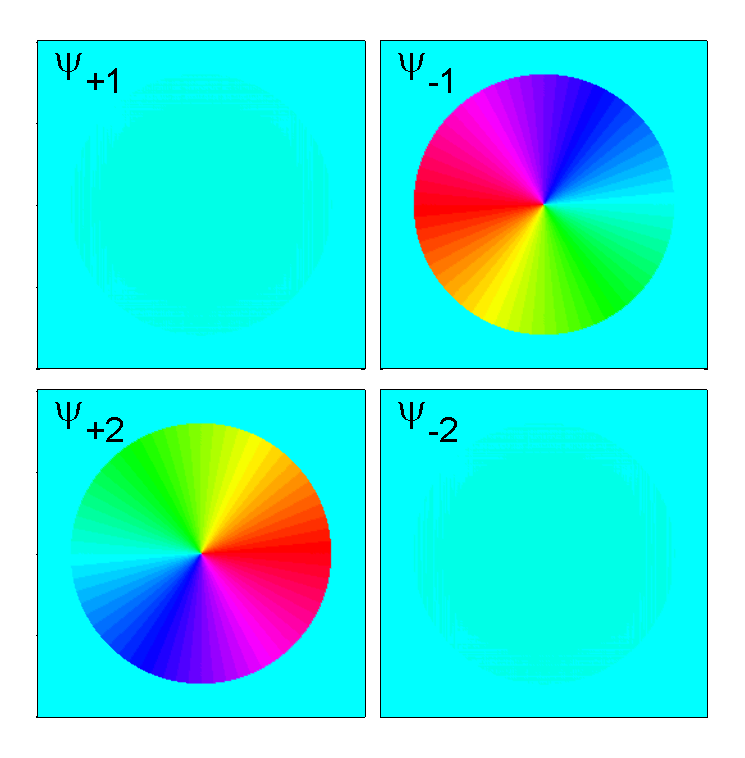}
  \end{minipage}

\caption{(Color online) Density and phase profiles of the exciton condensate with different topological defects. \textbf{Top:} $m_\sigma = (0,1,-1,0)$. \textbf{Second from top:} $m_\sigma = (1,-1,1,-1)$. \textbf{Second from bottom:} $m_\sigma = (1,1,1,1)$ and $k_\sigma = (1,-1,0,0)$. \textbf{Bottom:} $m_\sigma = (1,0,1,0)$ and $k_\sigma = (1,0,0,0)$. In the top three: $V_0 = 1$ and $W = -0.1$ $\mu$eV$\mu$m$^{-2}$. Bottom: $V_0 = 1$ and $W = 0.1$ $\mu$eV$\mu$m$^{-2}$. In all pictures: $u_0 = 1$ $\mu$eV$\mu$m$^{-2}$.}
\label{Fig3_x}
\end{figure}
Of main interest are configurations such as
$(m_{+2},m_{+1},m_{-1},m_{-2})$ = (0,1,-1,0), (1,1,1,1), (1,0,1,0)
satisfying Eq. (\ref{eq.bound1}). We observe that if Eq.
(\ref{eq.bound1}) is not satisfied, then there is no energy minimum
for a trapped state of the considered topological defect,
cylindrically symmetric or not. The real time propagation revealed
that if for example a stable solution of $m_\sigma = (0,1,-1,0)$ was
suddenly switched to $(0,1,1,0)$ by conjugating the $\Psi_{-1}$
component then the solution became immediately non-stationary and
the localized topological defect was destroyed.

The vortices with high topological charges, $|m_\sigma|
> 1$, were shown to be unstable in single component BECs depending
on interaction strength.\cite{Pu1999} This holds as well in our
case: single topological defects are no longer observed for
$|m_\sigma| > 1$ in the case when SOI is absent. This situation
changes, however, if SOI is taken into account as it will be
discussed in the next section.

One should note that in the four component BEC the term vortex
commonly applies when all components are rotating. The half vortex
pair corresponds to circular motion of two components in the same
direction, and half vortex-antivortex pair to two components with
opposite direction of rotation.

In Fig. \ref{Fig3_x} we show four cases of low energy solutions for
vortex topological defects in the four-component exciton condensate.
The top plots correspond to a half vortex-antivortex pair in $\Psi_{\pm
1}$. The second from the top corresponds to a basic vortex composed
of two half vortex-antivortex pairs in both bright and dark
components. The second from the bottom corresponds to a basic vortex
composed of two half vortex pairs in both bright and dark components
--- both with a $\pi$ phase difference. Bottom plots correspond to a half
vortex pair in $\Psi_{-1}$ and $\Psi_{+2}$ components. One can see
in the top and bottom lines that the vortex core stabilizes at a
greater healing length due to the other components trying to fill in
the density dips. The densities of bright and dark excitons try to
complement each other, staying close to the Thomas-Fermi profile.

The existence of a low energy solution with vortices is determined
by the last term in Eqs. (\ref{eq.gp5})-(\ref{eq.gp8}) and the
kinetic energy term. This can be seen from analysis of the
Hamiltonian density (\ref{eq.hamdens}). In the case of $W=0$ there
is a competition between the kinetic energy term in the total
Hamiltonian and the interaction energy term
\[
\mathcal{H}_{\text{int}}=\frac{V_0}{2}\left(|\Psi_{+2}|^2+|\Psi_{+1}|^2+|\Psi_{-1}|^2+|\Psi_{-2}|^2\right)^2.
 \]
If interactions are weak then it will be energetically favorable to
transfer intensity from a component with a vortex to a component
without one, since a component with a vortex has higher kinetic
energy. For this reason, there may be no minimal energy states with
vortices in two components only --- numerical calculations give
instead a depletion of components containing vortices in favor of
those without vortices. We can then expect that the only possible
stable states with non-zero topological charges (in components with
non-zero intensity) contain vortices in all components.

One should keep in mind that while the kinetic energy contribution
can be reduced by transferring intensity to a component without a
vortex, this may increase the potential energy due to interactions.
The term proportional to $V_0$ in the Hamiltonian can be reduced if
the spatial overlap of the intensity distribution of components is
reduced. Thus the $V_0$ term favors formation of vortices, but it
must be strong enough to overcome the corresponding increase of
kinetic energy for the states with vortices in two components only.

In the case $W\neq0$ and $(m_{+2},m_{+1},m_{-1},m_{-2})=(0,1,-1,0)$
the wave functions can be written:
\[
\Psi_{+2}=U(r)e^{i\phi_{+2}}, \quad \Psi_{+1}=V(r)e^{i\theta}e^{i\phi_{+1}}
\]
\[
\Psi_{-1}=V(r)e^{-i\theta}e^{i\phi_{-1}}, \quad \Psi_{-2}=U(r)e^{i\phi_{-2}}
\]
where $U(r)$ and $V(r)$ are real functions. The $W$ dependent part of the Hamiltonian can then be written as
\begin{equation}
\mathcal{H}_W=W\left[2U^2V^2\cos{\left(\Delta \phi \right)} -U^4-V^4\right],
\end{equation}
where $\Delta \phi = \phi_{+2}+\phi_{-2}-\phi_{+1}-\phi_{-1}$.

In the case that $W>0$, the phases can be chosen to minimize the
Hamiltonian to
$\mathcal{H}_W=-W\left(U^2+V^2\right)^2= - W \left(|\Psi_{+2}|^2+|\Psi_{+1}|^2+|\Psi_{-1}|^2+|\Psi_{-2}|^2\right)^2$.
That is, the $W$ term has the same form as the $V_0$ term.
Consequently, the same arguments as considered in the $W=0$ case
apply: if the strength of the interaction $V_0-W$ is unable to
overcome the kinetic energy term, then a state with vortices in two
components only (with non-zero intensity) is not stable.

In the case that $W<0$, the phase can be chosen to minimize the
Hamiltonian to $\mathcal{H}_W=-W\left(U^2-V^2\right)^2$ (a positive
quantity since $W<0$). This term favors states with vortices in two
components (see Fig. \ref{Fig3_x}, top case) since it is minimized
if all components stay populated.

Note that it is not possible to say definitively whether vortices
will or will not be stable for the cases where $W$ is positive or
negative without use of numerical calculation because of the tricky
interplay between potential and kinetic energy terms.

In the case of $W\neq0$ and $(m_{+2},m_{+1},m_{-1},m_{-2})=(1,0,1,0)$ the wave functions can be written:
\[
\Psi_{+2}=V(r)e^{i\theta}e^{i\phi_{+2}}, \quad \Psi_{+1}=U(r)e^{i\phi_{-1}}
\]
\[
\Psi_{-1}=V(r)e^{i\theta}e^{i\phi_{+1}}, \quad \Psi_{-2}=U(r)e^{i\phi_{-2}}.
\]
The $W$ dependent part of the Hamiltonian density is
\begin{equation}
\mathcal{H}_W=2WU^2V^2\left[\cos{\left(\Delta \phi \right)}-1\right].
\end{equation}
In the case $W>0$, the phases can be chosen to minimize the
Hamiltonian to $\mathcal{H}_W=-4WU^2V^2$. This term may stabilize
the state, since it provides a reduction of the energy when all
components are populated (see Fig. \ref{Fig3_x}, bottom case); if
one component is depleted then this term can no longer contribute to
minimization of the energy.

In the case $W<0$, the phases can be chosen to minimize the
Hamiltonian to $\mathcal{H}_W=0$. In this case we recover the result
for the $W=0$ case. While it is not possible to say definitively
whether vortices will be stable for the $W>0$, we can say that for
$W<0$ they are unstable if they are also unstable for the $W=0$
case.

\section{Cylindrically symmetric ground state solutions under Spin-Orbit interaction}
When SOI of Rashba or/and Dresselhaus type is included in the
Hamiltonian, the analysis of low energy state solutions becomes more
tricky. Prior studies in the field of atomic condensates revealed a
plethora of phenomena emerging due to spin-orbit
coupling.\cite{Lin2011,Galitski2013} Indirect exciton condensates
can be expected to show also a great variety in possible low energy
solutions with phase separation between components and density
modulations.
\begin{figure}[h!]
\centering

  \begin{minipage}[b]{0.41\linewidth}
    \centering
    \includegraphics[width=\linewidth]{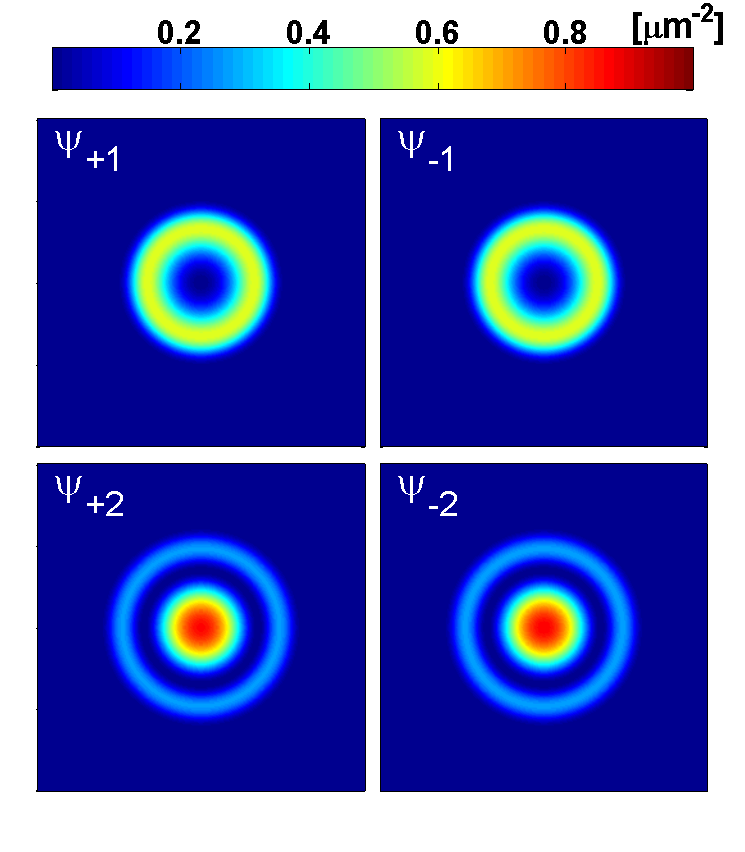}
  \end{minipage}
  \hspace{0cm}
  \begin{minipage}[b]{0.41\linewidth}
    \centering
    \includegraphics[width=\linewidth]{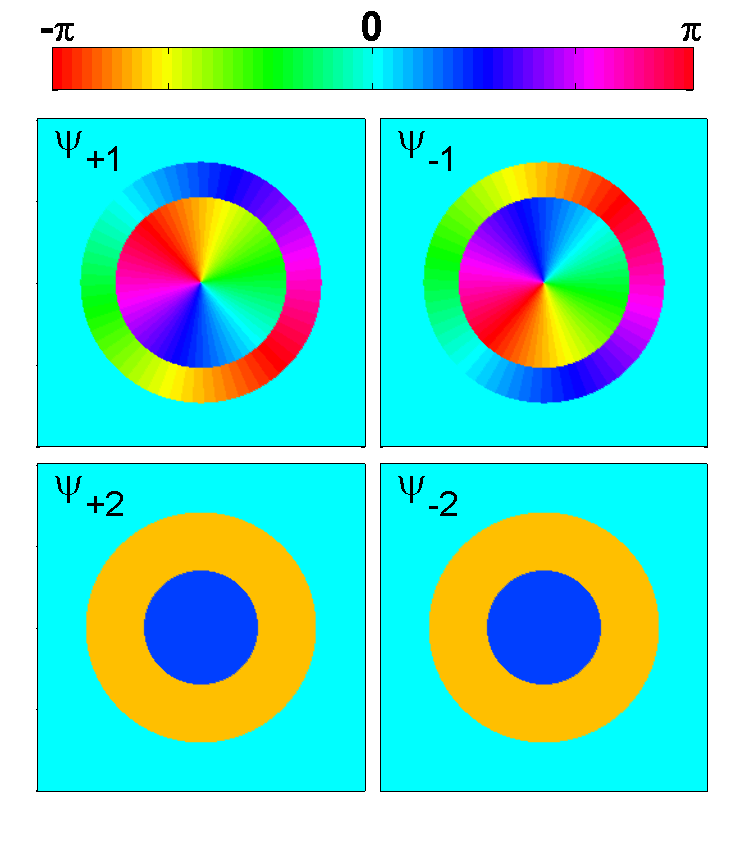}
  \end{minipage}

\centering
  \begin{minipage}[b]{0.41\linewidth}
    \centering
    \includegraphics[width=\linewidth]{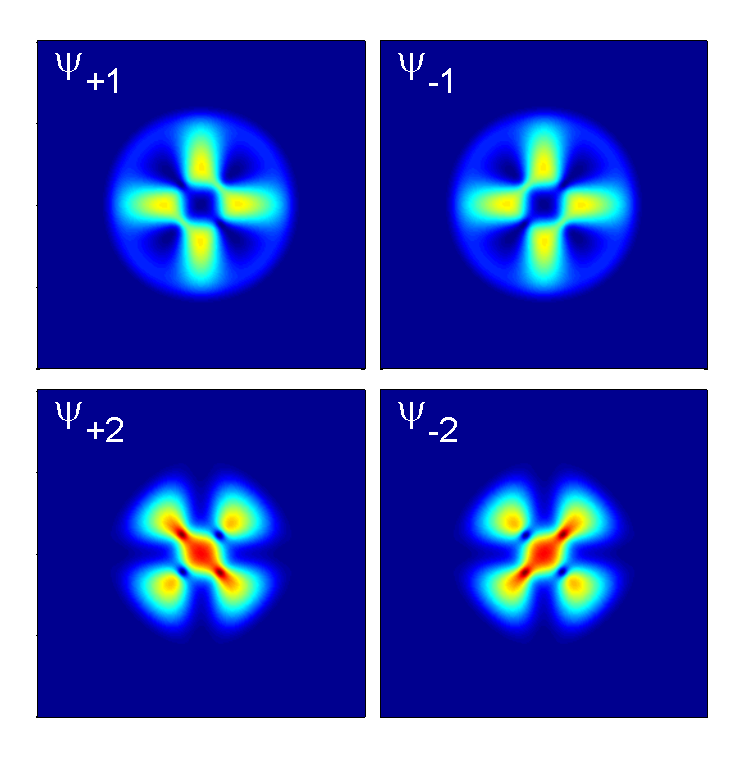}
  \end{minipage}
  \hspace{0cm}
  \begin{minipage}[b]{0.41\linewidth}
    \centering
    \includegraphics[width=\linewidth]{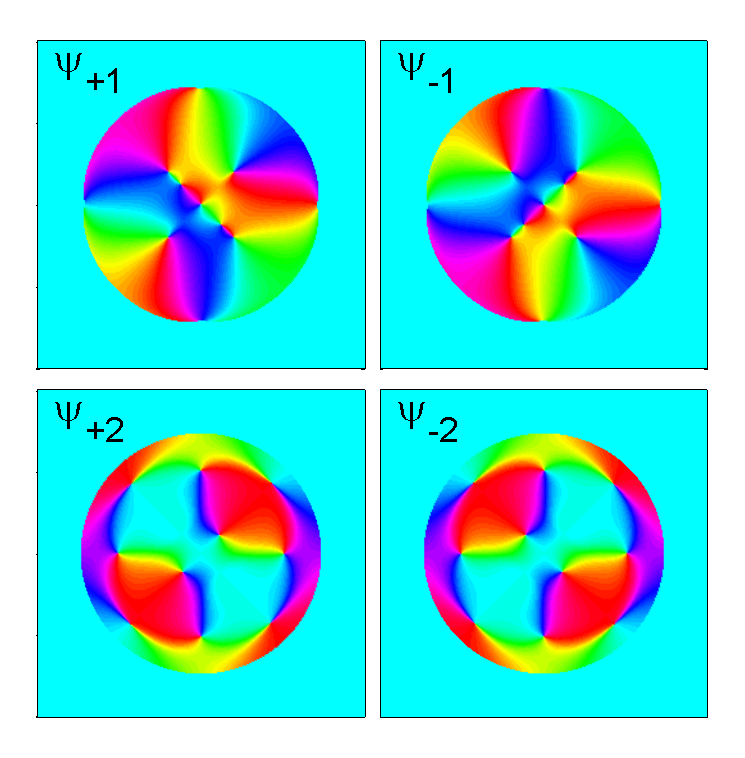}
  \end{minipage}

\centering
  \begin{minipage}[b]{0.41\linewidth}
    \centering
    \includegraphics[width=\linewidth]{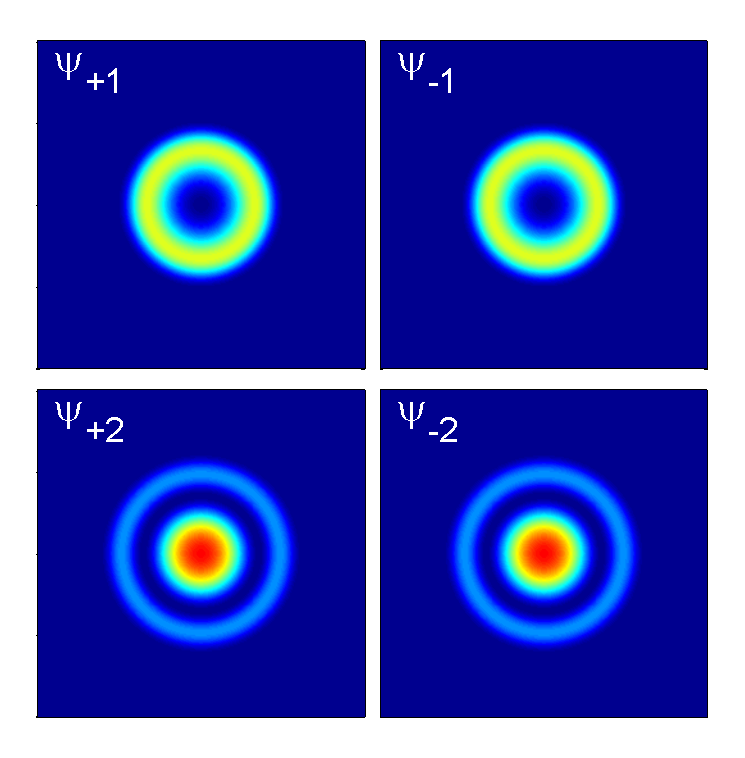}
  \end{minipage}
  \hspace{0cm}
  \begin{minipage}[b]{0.41\linewidth}
    \centering
    \includegraphics[width=\linewidth]{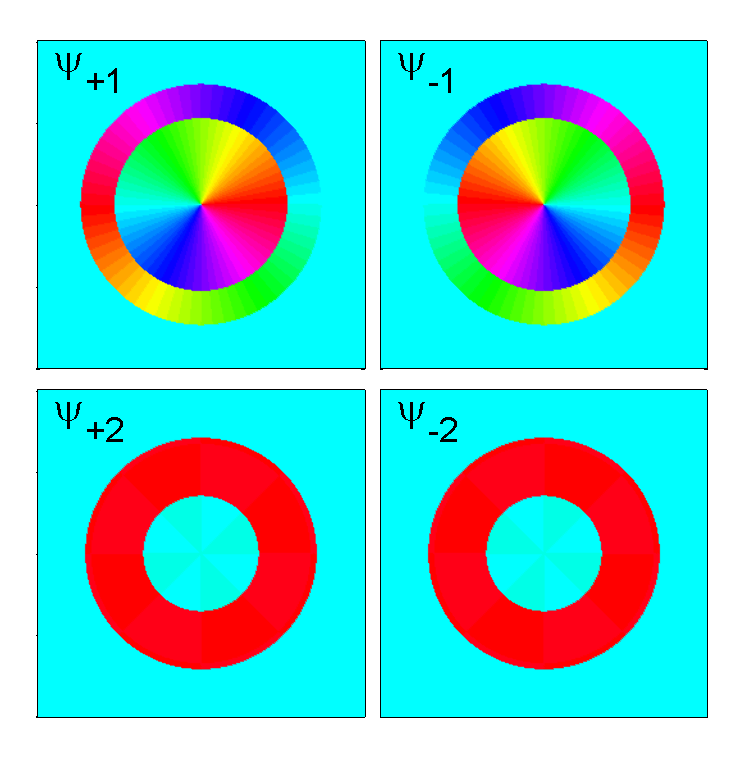}
  \end{minipage}

\centering
  \begin{minipage}[b]{0.41\linewidth}
    \centering
    \includegraphics[width=\linewidth]{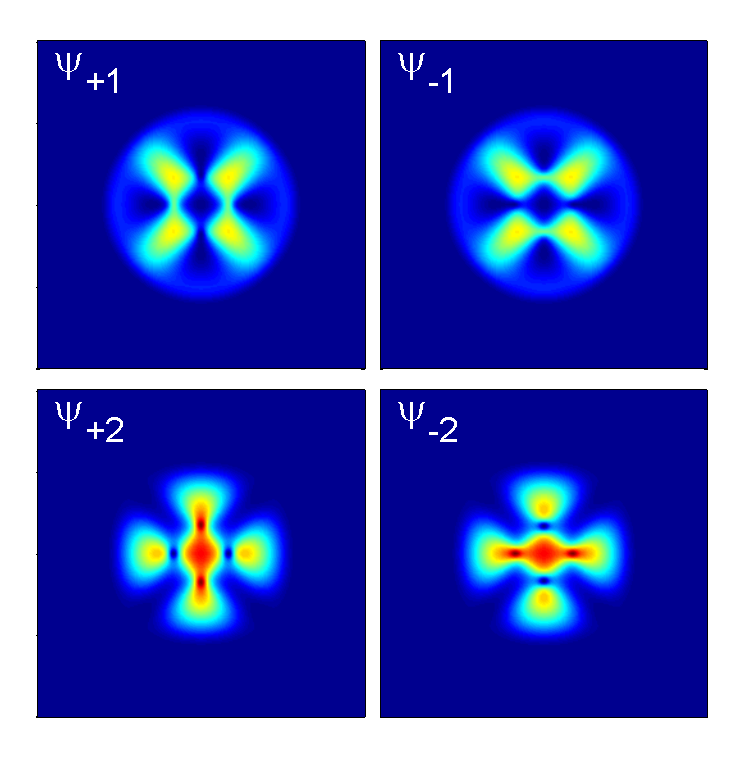}
  \end{minipage}
  \hspace{0cm}
  \begin{minipage}[b]{0.41\linewidth}
    \centering
    \includegraphics[width=\linewidth]{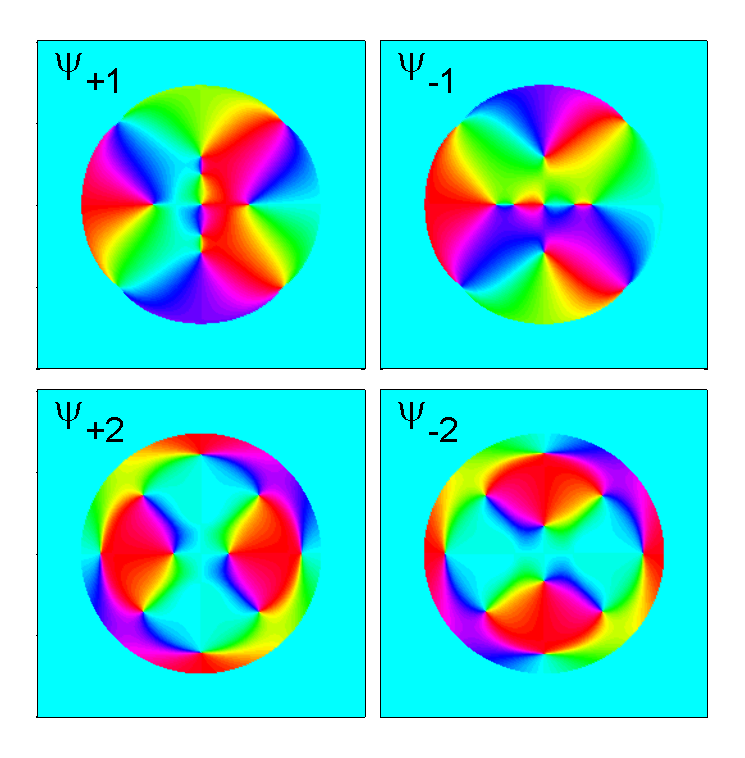}
  \end{minipage}
\caption{(Color online) The difference between Dresselhaus and
Rashba SOI displayed. Density and phase profiles of the condensate
components with different vortex defects as initial condition. In
the top two cases: $\beta =1$ $\mu$eV$\mu$m and $\alpha = 0$. In the
lower two cases: $\beta =0$ and $\alpha = 1$ $\mu$eV$\mu$m.
\textbf{Top:} $\beta =1$ $\mu$eV$\mu$m and $\alpha = 0$, initial configuration corresponds to $m_\sigma =
(0,-1,1,0)$. The bound (\ref{eq.bound2}) is satisfied and
cylindrically symmetric vortex type solution is obtained.
\textbf{Second from top:} $\beta =1$ $\mu$eV$\mu$m and $\alpha = 0$, $m_\sigma = (0,1,-1,0)$. The bound (\ref{eq.bound2}) is not satisfied, and as a result warped vortex corresponding to $m_\sigma = (+2,+3,-3,-2)$ is formed in a stationary regime. \textbf{Second from bottom:}  $\beta =0$ and $\alpha = 1$ $\mu$eV$\mu$m, $m_\sigma = (0,1,-1,0)$. The bound (\ref{eq.bound3}) is satisfied and cylindrically symmetric vortex type solution is obtained. \textbf{Bottom:} $\beta =0$ and $\alpha = 1$ $\mu$eV$\mu$m, initial configuration corresponds to $m_\sigma = (0,-1,1,0)$. The bound (\ref{eq.bound3}) is not satisfied, and as a result warped vortex corresponding to $m_\sigma = (-2,-3,+3,+2)$ is formed in stationary regime. In all pictures: $k_\sigma=(0,0,0,0)$, $V_0 = 22$ $\mu$eV$\mu$m$^{-2}$, $W = 2$ $\mu$eV$\mu$m$^{-2}$ and $u_0 = 1$ $\mu$eV$\mu$m$^{-2}$. }
\label{Fig4_x}
\end{figure}

Let us first analyze the possibility of the cylindrically symmetric
solutions. Using the ansatz $\Psi_\sigma=R_\sigma(r)e^{im_\sigma
\theta}$ in Eqs. (\ref{eq.gp1})-(\ref{eq.gp4}) we find that if only
Dresselhaus SOI is present, the winding numbers should satisfy the
following bound [in addition to those given by Eq.
(\ref{eq.bound1})]:
\begin{align} \label{eq.bound2}
&m_{+2} = 1 + n, \  \quad m_{+1} = n, \\ \notag
&m_{-1} = 1 + m, \  \quad m_{-2} = m.
\end{align}
On the other hand, if only Rashba SOI is present the bound is:
\begin{align} \label{eq.bound3}
&m_{+2} = n, \  \quad m_{+1} = 1+n, \\ \notag
&m_{-1} = m, \  \quad m_{-2} = 1+m,
\end{align}
where $n$ and $m$ are integer numbers.
\begin{figure}[t!]
\centering
  \begin{minipage}[b]{0.48\linewidth}
    \centering
    \includegraphics[width=\linewidth]{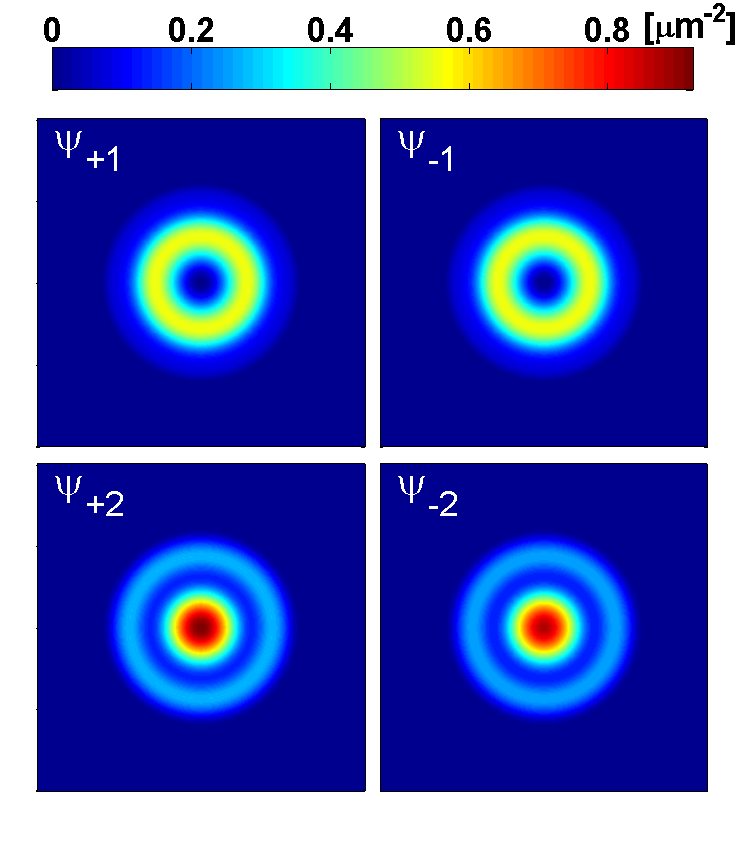}
  \end{minipage}
  \hspace{0cm}
  \begin{minipage}[b]{0.48\linewidth}
    \centering
    \includegraphics[width=\linewidth]{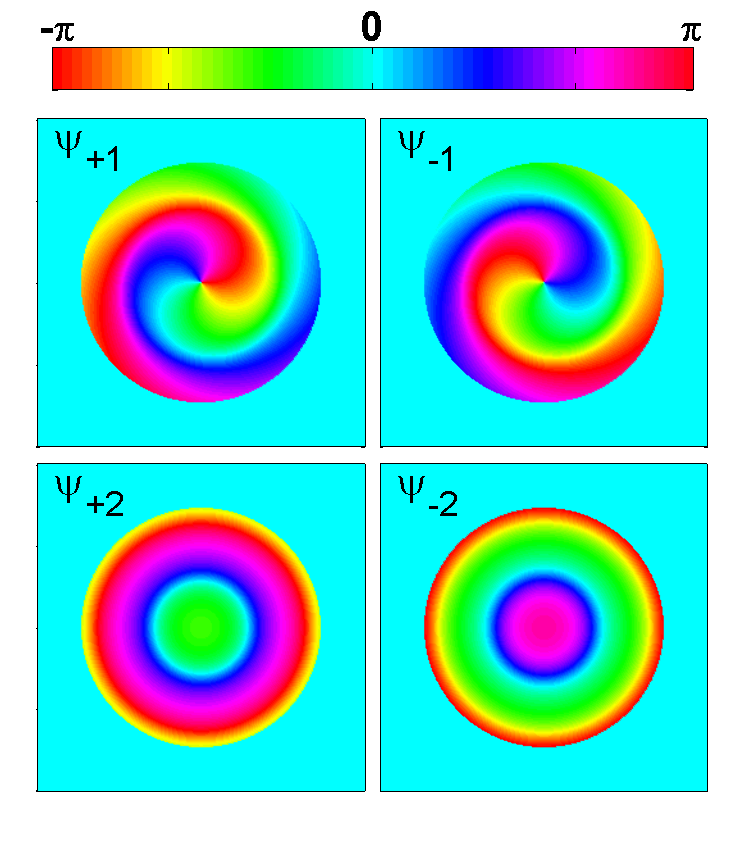}
  \end{minipage}

\centering
    \includegraphics[width=\linewidth]{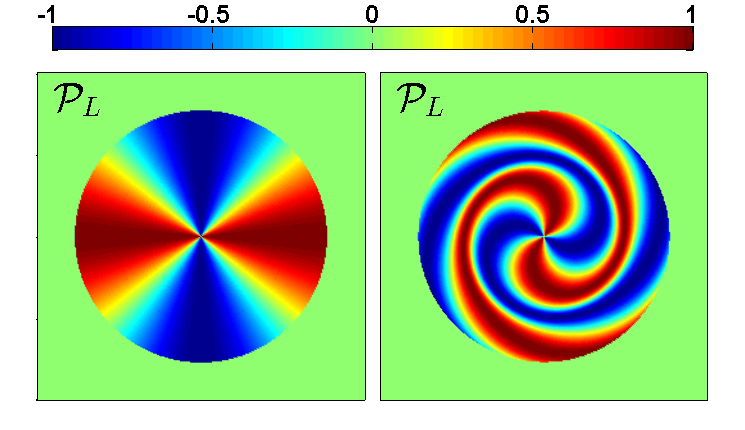}

\caption{(Color online) \textbf{Top:} Density and phase profiles of the exciton condensate components for only Dresselhaus SOI. $m_\sigma = (0,-1,1,0)$, $k_\sigma = (0,1,0,0)$ and $W = 2~\mu$eV$\mu$m$^{-2}$, $V_0 = 28$ $\mu$eV$\mu$m$^{-2}$, $\beta = 1~\mu$eV$\mu$m and $u_0 = 1$ $\mu$eV$\mu$m$^{-2}$. \textbf{Bottom:} Linear polarization of cylindrically symmetric cases for Dresselhaus SOI only with $m_\sigma = (0,-1,1,0)$ initial condition. The \textbf{right} panel was calculated for  $k_\sigma = (0,0,0,0)$, and corresponds to the top panel in Fig. \ref{Fig4_x}. For the \textbf{left} panel we set $k_\sigma = (0,1,0,0)$, corresponding to the top panel in this figure.}
\label{Fig51_x}
\end{figure}

We limit our consideration in this section of the paper to three
types of cylindrical vortex configurations for SOI of either
Dresselhaus or Rashba type: $m_\sigma = (0,1,-1,0), \ (0,-1,1,0)$,
and $(1,0,1,0)$. We show that in the case where the bounds
(\ref{eq.bound2}) or (\ref{eq.bound3}) are not satisfied the initial
topological charge is not preserved but instead the
configuration of \emph{warped vortices} with winding numbers greater than one, $m_\sigma =
(\pm2,\pm3,\mp3,\mp2)$,\cite{RuboComment} or density modulated stripe phase with no vorticity is established, depending on the initial conditions. 
\begin{figure}[t!]
\centering
  \begin{minipage}[b]{0.48\linewidth}
    \centering
    \includegraphics[width=\linewidth]{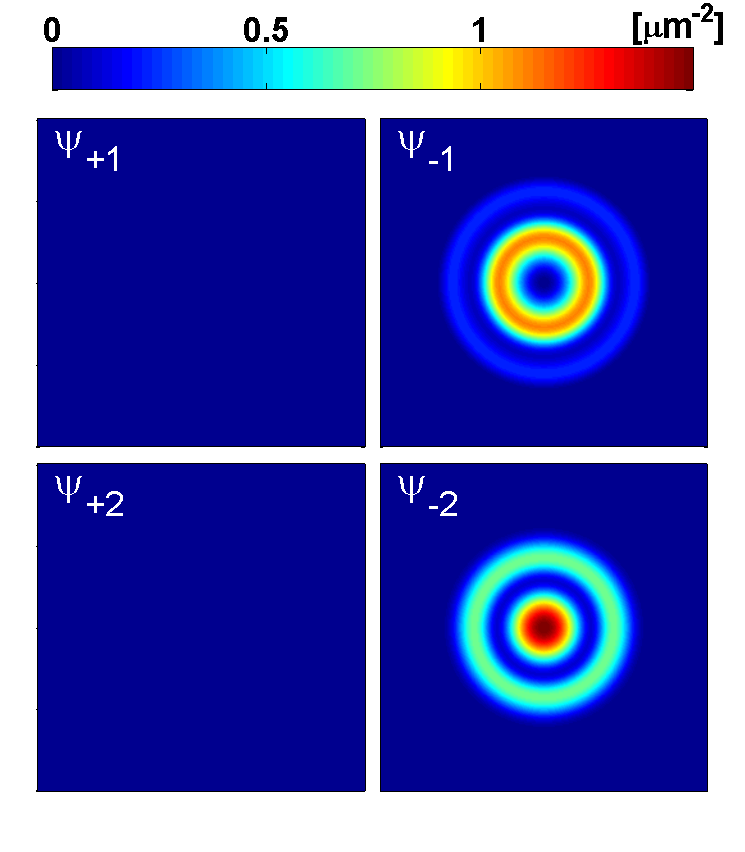}
  \end{minipage}
  \hspace{0cm}
  \begin{minipage}[b]{0.48\linewidth}
    \centering
    \includegraphics[width=\linewidth]{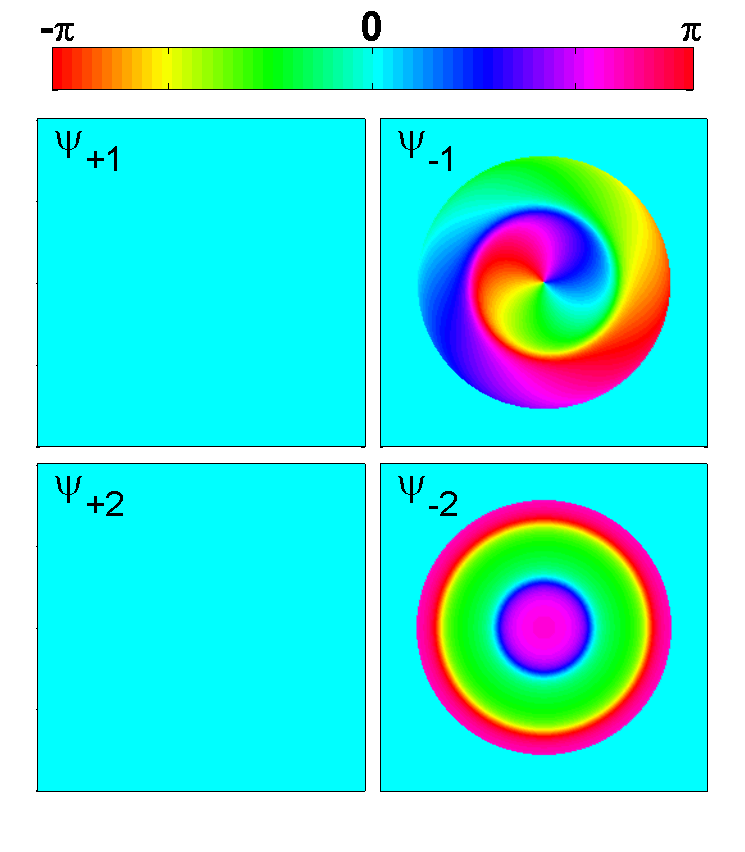}
  \end{minipage}

\caption{(Color online) Density and phase profiles of the exciton condensate
components for only Dresselhaus SOI. $m_\sigma = (0,-1,1,0)$, $k_\sigma = (0,0,0,0)$, $W = -2~\mu$eV$\mu$m$^{-2}$, $V_0 = 28$ $\mu$eV$\mu$m$^{-2}$, $\beta = 1~\mu$eV$\mu$m and $u_0 = 1$ $\mu$eV$\mu$m$^{-2}$.}
\label{Fig512_x}
\end{figure}
\begin{figure}[t!]
\centering
  \begin{minipage}[b]{0.48\linewidth}
    \centering
    \includegraphics[width=\linewidth]{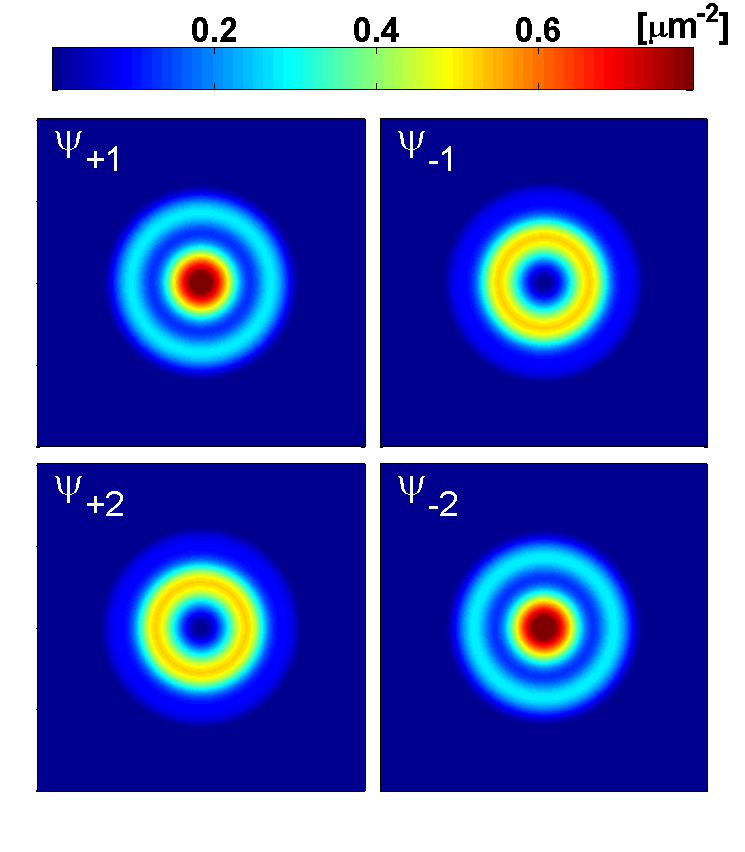}
  \end{minipage}
  \hspace{0cm}
  \begin{minipage}[b]{0.48\linewidth}
    \centering
    \includegraphics[width=\linewidth]{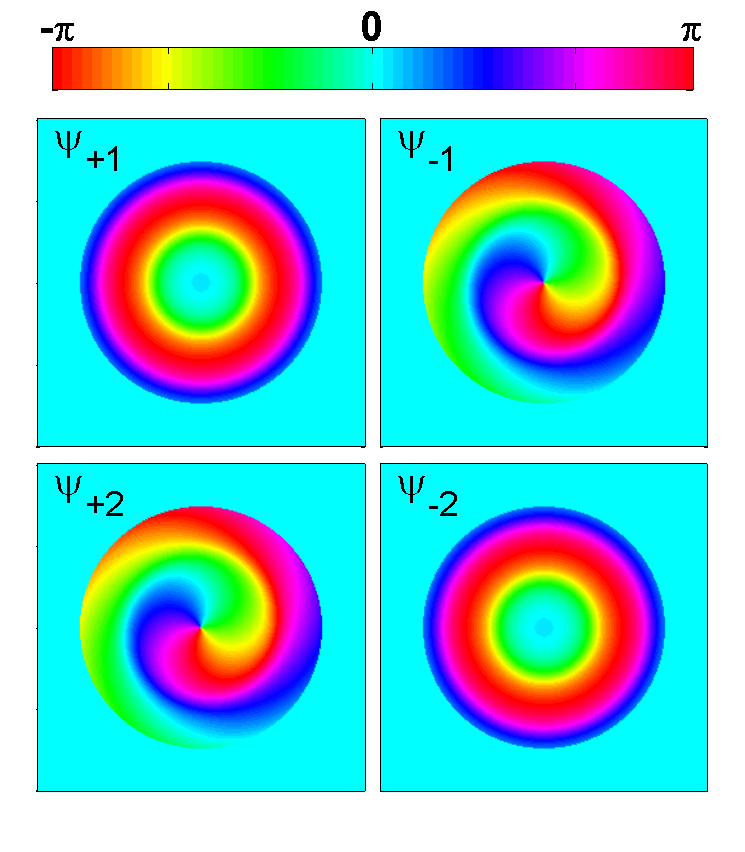}
  \end{minipage}
\caption{(Color online) Density and phase profiles of the exciton condensate half vortex pair for only Dresselhaus SOI. $m_\sigma = (1,0,1,0)$, $k_\sigma = (0,0,0,0)$, $V_0 = 28$ $\mu$eV$\mu$m$^{-2}$, $W = \pm2~\mu$eV$\mu$m$^{-2}$, $\beta = 1~\mu$eV$\mu$m and $u_0 = 1$ $\mu$eV$\mu$m$^{-2}$.}
\label{Fig52_x}
\end{figure}

For Dresselhaus SOI only the configurations $m_\sigma = (0,-1,1,0)$
and $(1,0,1,0)$ satisfy Eq. (\ref{eq.bound2}), and we observe
formation of the cylindrically symmetric vortices (Fig. \ref{Fig4_x}
top and Fig. \ref{Fig52_x}), whereas $m_\sigma = (0,1,-1,0)$ does
not satisfy the bound, the cylindrical symmetry is no longer present,
and configuration with higher winding numbers $m_\sigma =
(+2,+3,-3,-2)$ is formed (Fig. \ref{Fig4_x}, second from top). The
similar behavior can be observed for the case of the Rashba SOI but
this time the cylindrical symmetry is manifested for $m_\sigma =
(0,1,-1,0)$ and $(0,1,0,1)$ configurations.

Fig. \ref{Fig4_x} shows stable solutions for the half
vortex-antivortex configurations $m_\sigma = (0,1,-1,0)$, and
$(0,-1,1,0)$. The top two panels correspond to the case when only
Dresselhaus SOI is present and the bottom two for the case when only
Rashba SOI is present. Inspecting the phase profiles (top and second
from bottom panel) reveals that phases of the components are
different for the cases of Dresselhaus and Rashba SOI: there is a
$3\pi/4$ phase difference in $\Psi_{+1}$, $-\pi/4$ difference in
$\Psi_{-1}$ and $\pi/4$ difference in $\Psi_{\pm 2}$ amplitudes if
one changes Dresselhaus SOI to Rashba. These phase differences
result in the $\pi/2$ rotation of the pattern of the linear
polarization degree calculated as
\begin{equation}
\mathcal{P}_L = \frac{\Psi_{+1}^* \Psi_{-1} + \Psi_{-1}^* \Psi_{+1}
}{|\Psi_{+1}|^2 + |\Psi_{-1}|^2 } \propto \cos{(2 \theta)}
\end{equation}
if one switches from Rashba to Dresselhaus SOI. The result is
expectable, as the operator describing Rashba SOI can be obtained
from the operator describing Dresselhaus SOI by switching $K_x$ to
$K_y$ and vice versa [see Eq. (\ref{eq.ksoi})].
\begin{figure}[t!]
\centering
  \begin{minipage}[b]{0.48\linewidth}
    \centering
    \includegraphics[width=\linewidth]{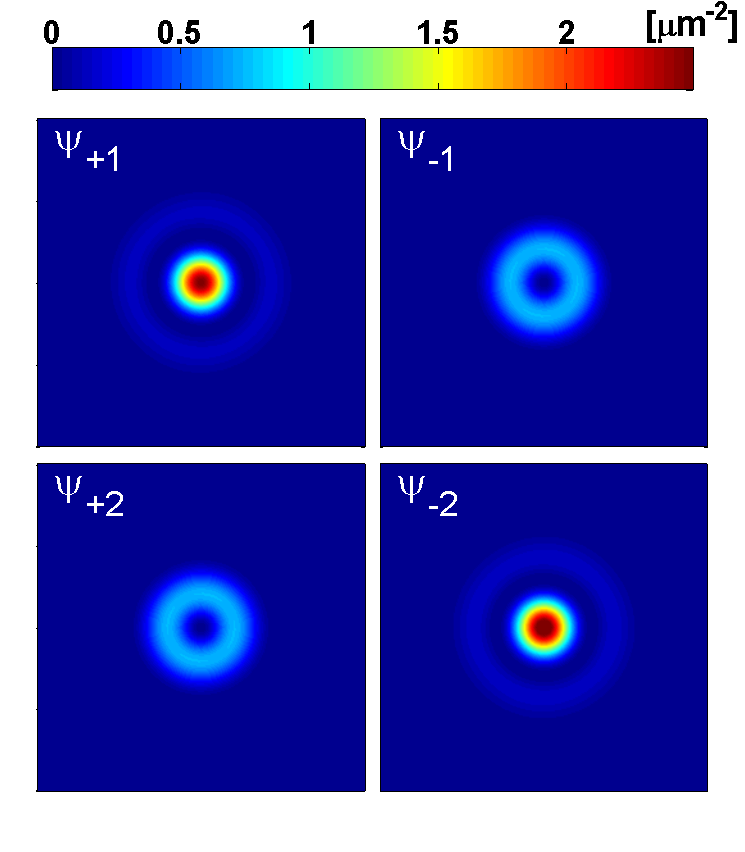}
  \end{minipage}
  \hspace{0cm}
  \begin{minipage}[b]{0.48\linewidth}
    \centering
    \includegraphics[width=\linewidth]{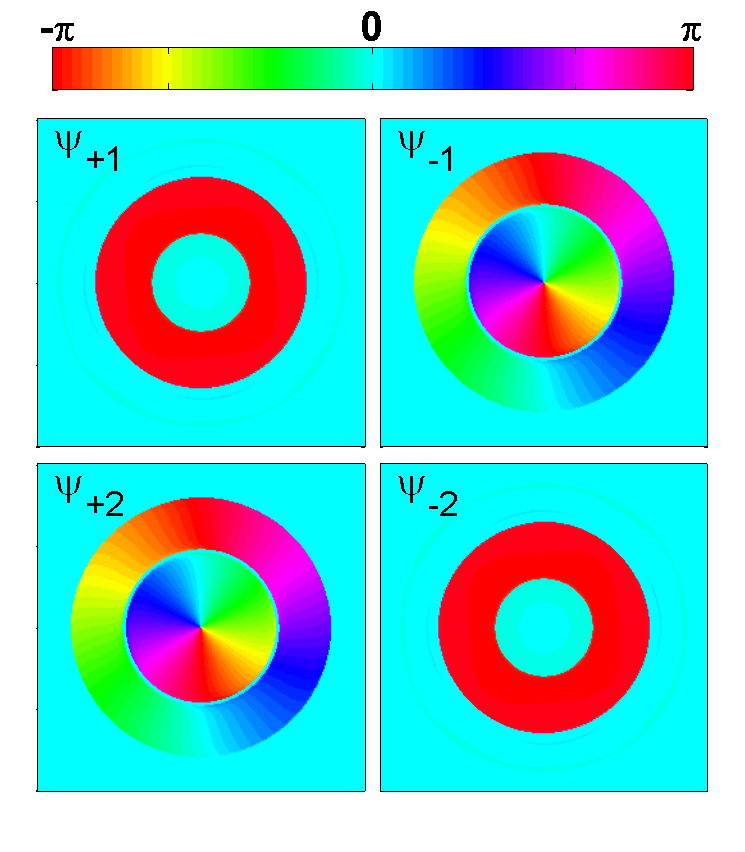}
  \end{minipage}

\centering
  \begin{minipage}[b]{0.48\linewidth}
    \centering
    \includegraphics[width=\linewidth]{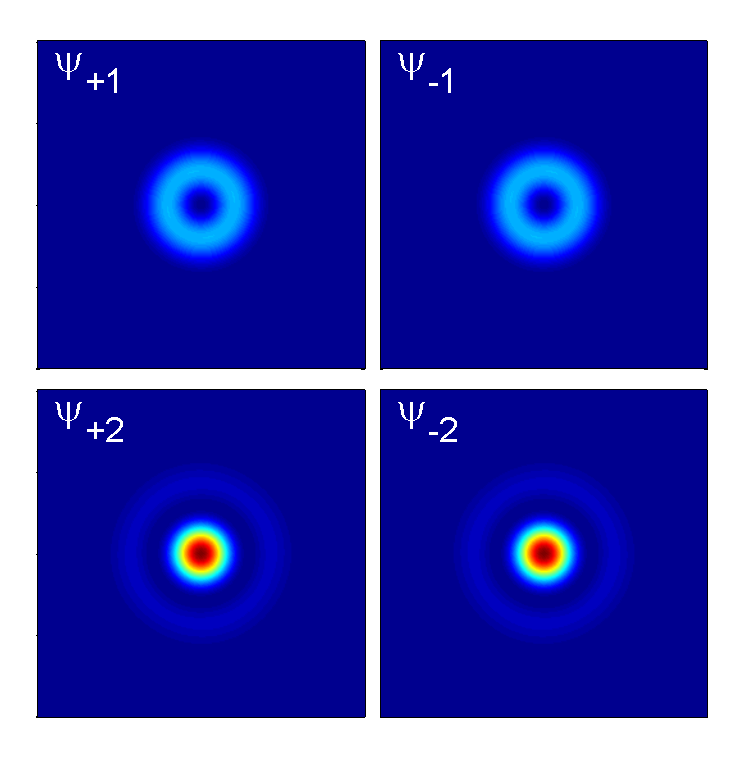}
  \end{minipage}
  \hspace{0cm}
  \begin{minipage}[b]{0.48\linewidth}
    \centering
    \includegraphics[width=\linewidth]{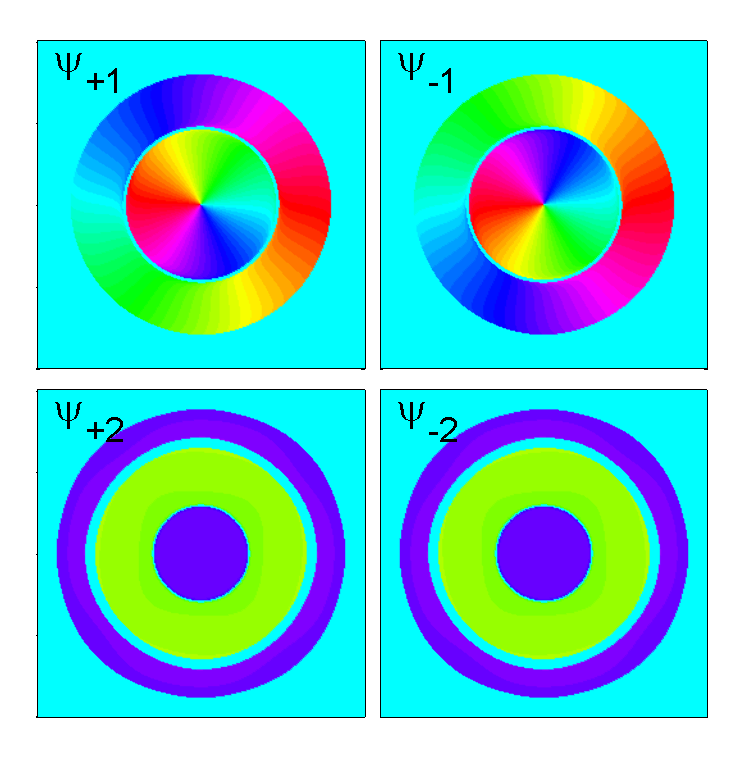}
  \end{minipage}

\centering
  \begin{minipage}[b]{0.48\linewidth}
    \centering
    \includegraphics[width=\linewidth]{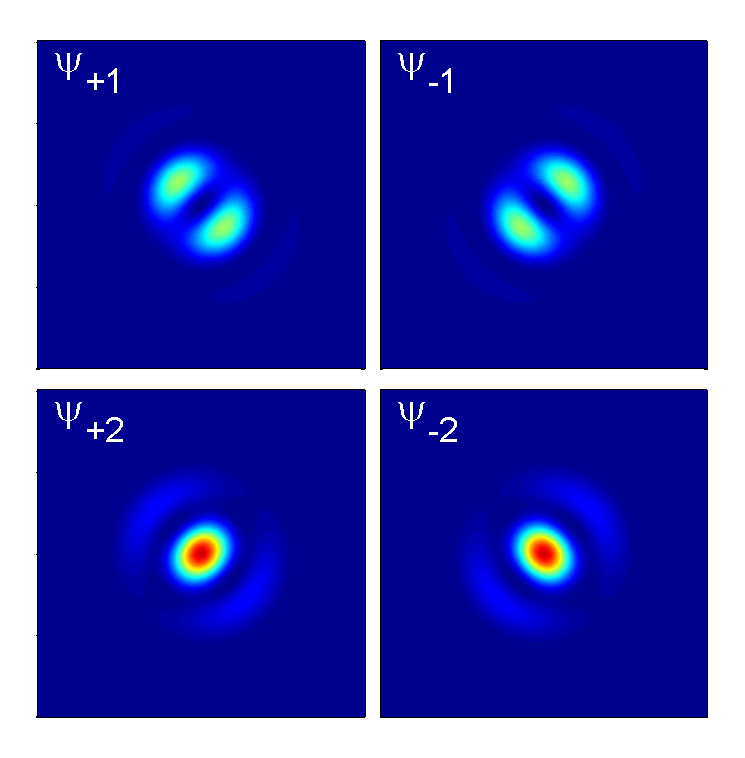}
  \end{minipage}
  \hspace{0cm}
  \begin{minipage}[b]{0.48\linewidth}
    \centering
    \includegraphics[width=\linewidth]{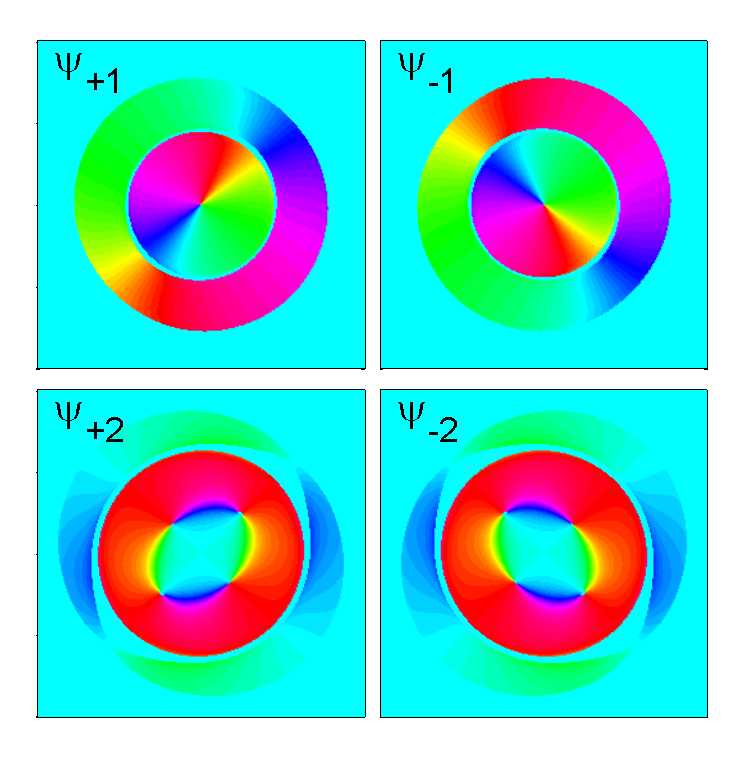}
  \end{minipage}
\caption{(Color online) Density and phase profiles of the exciton condensate
components for only Dresselhaus SOI and small nonlinear parameters.
\textbf{Top:} $m_\sigma = (1,0,1,0)$. \textbf{Middle:} $m_\sigma =
(0,-1,1,0)$. \textbf{Bottom:} $m_\sigma = (0,1,-1,0)$. In all
pictures the parameters were  $\beta = 1$ $\mu$eV$\mu$m, $V_0 = 2.8$ neV$\mu$m$^{-2}$, $W = \pm0.2$ neV$\mu$m$^{-2}$ and $u_0 = 1$ neV$\mu$m$^{-2}$.} \label{Fig6_x}
\end{figure}

We observe that in the cylindrically symmetric case there are two
alternative configurations of the vortex corresponding to the same
combination of the winding numbers. One of them is demonstrated in
Fig. 3 and corresponds to the case when the phase of macroscopic
wavefunction depends only on the angle $\phi$. This solution is
obtained if all $k_\sigma$ are put to zero. However, if one
introduces phase difference between the condensate components chosen
as an initial condition (i.e. $k_\sigma \neq 0$), another type of
the vortex solution corresponding to the spiral phase pattern is
obtained [see Fig. \ref{Fig51_x}, top]. The topological charges of
both solutions are the same, and to distinguish between them one
needs to analyze their linear polarization patterns shown in Fig.
\ref{Fig51_x} (bottom). As one can see, they are radically different, being four leaf in one case and gammadion in the other.

Also we note that the sign of exchange interaction $W$ affects the possible states of stable topological defects. To illustrate its role, we focus on a configuration $m_\sigma = (0,-1,1,0)$ (same as in top panel in Fig. \ref{Fig4_x}) and set the parameters to: $\beta = 1$ $\mu$eV$\mu$m, $\alpha = 0$ and $W = -2~\mu$eV$\mu$m$^{-2}$. We observe a half vortex in a condensate half depleted with a spiral phase pattern resulting from negative $W$ [see Fig. \ref{Fig512_x}]. The results are clearly different from those shown in Fig. \ref{Fig4_x} corresponding to opposite sign of the exchange interaction, $W = +2$ $\mu$eV$\mu$m$^{-2}$.

In Fig. \ref{Fig52_x} we show a half vortex pair solution with $m_\sigma = (1,0,1,0)$ for only Dresselhaus SOI [in case of only Rashba it would be $m_\sigma = (0,1,0,1)$]. The solution remained the same for both signs of the mixing parameter $W$ and was lost when $k_\sigma \neq 0$.

We also investigated the ground state vortex solutions for the case
when the nonlinearities are very weak and the impact of the SOI
terms becomes dominant $(V_0,W \ll \beta, \alpha)$. Fig.
\ref{Fig6_x} illustrates the case when only Dresselhaus SOI is
present. If the bound (\ref{eq.bound2}) is satisfied the solutions
have cylindrical symmetry (top two panels). For the case $m_\sigma =
(0,1,-1,0)$ the solution is non-symmetric and resembles a hybrid of
a warped vortex solution and a striped phase. In this weakly
nonlinear limit the sign of $W$ becomes irrelevant, and the same
patterns were observed for $W = \pm 0.2$ neV$\mu$m$^{-2}$.
\begin{figure}[t!]
\centering
  \begin{minipage}[b]{0.48\linewidth}
    \centering
    \includegraphics[width=\linewidth]{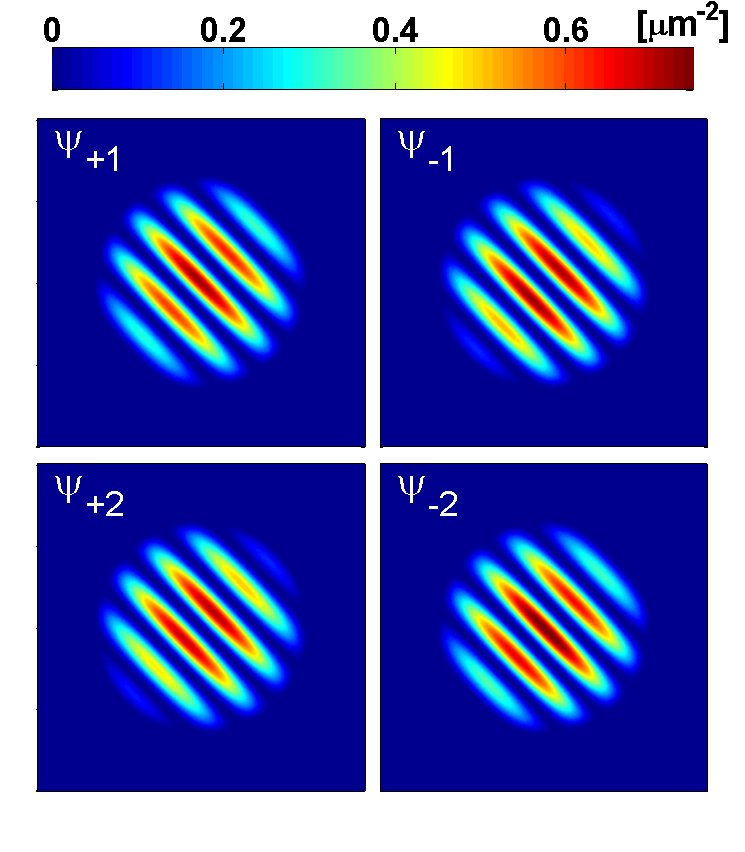}
  \end{minipage}
  \hspace{0cm}
  \begin{minipage}[b]{0.48\linewidth}
    \centering
    \includegraphics[width=\linewidth]{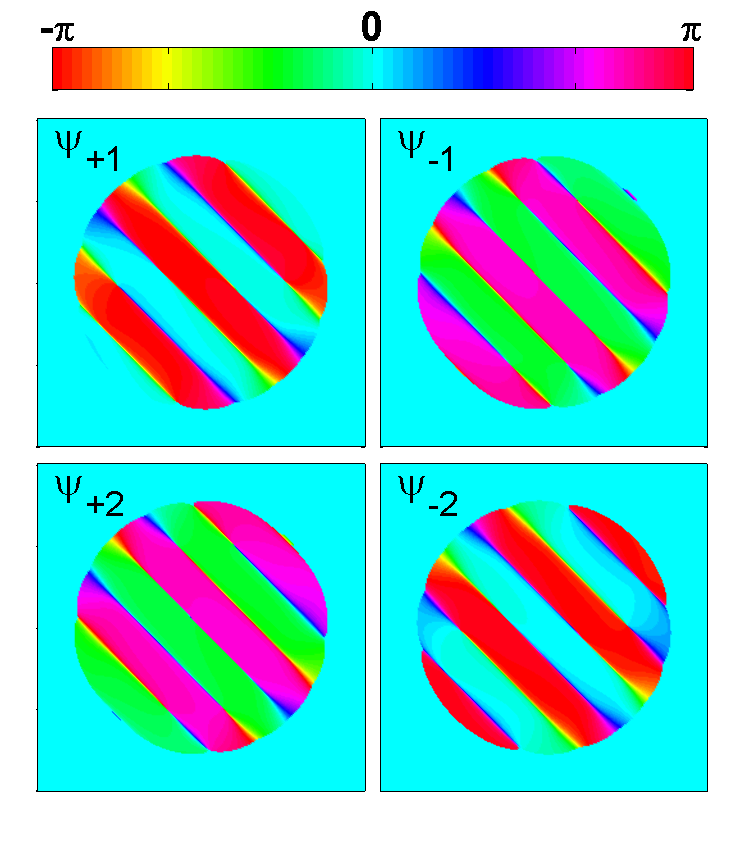}
  \end{minipage}
\caption{(Color online) Density and phase profile of condensate components for $m_\sigma = (0,0,0,0)$ and $k_\sigma = (0,1,0,0)$, the parameters were $\beta = 1$ $\mu$eV$\mu$m, $\alpha/\beta = 1/2$, $V_0 = 28$ $\mu$eV$\mu$m$^{-2}$ and $W = -2~\mu$eV$\mu$m$^{-2}$. }
\label{Fig7_x}
\end{figure}
\section{Presence of both Dresselhaus and Rashba SOI}
When both $\alpha \neq 0$ and $\beta \neq 0$ the single particle
spectrum becomes anisotropic. Different from the cases $\alpha=0$ or
$\beta=0$ the minima of the energy of non-interacting particles
correspond not to a circle of constant radius in the reciprocal
space, but to the two fixed points situated along $K_x$-$K_y$
diagonal,\cite{SpintronicsReview}
\begin{equation} \label{eq.qval}
\v{K}_0 = \pm \frac{\chi m_X(\alpha +
\beta)}{\hbar^2}\frac{(\v{e}_x+\v{e}_y)}{\sqrt{2}}.
\end{equation}
One can thus expect formation of a striped ground state
corresponding to the spatial modulation of the density $(e^{i
\v{K}_0 \cdot \v{r}} + e^{-i \v{K}_0 \cdot \v{r}}) = 2
\cos{(\v{K}_0 \cdot \v{r})}$.
This is indeed the case as can be seen in Fig. \ref{Fig7_x}. As the
ground state of the condensate reveals spatial anisotropy, no
cylindrically symmetric vortex solutions can be expected to appear
in this case.

The stability of the vortex-type versus striped phase solutions
depends on the ratio $\alpha/\beta$. Fixing the parameters
describing nonlinearities as $V_0 = 28~\mu$eV$\mu$m$^{-2}$ and $W =
\pm 2$ $\mu$eV$\mu$m$^{-2}$, our numerical analysis shows that for
$\alpha/\beta \sim 10^{-3}$ the vortex type solutions shown in Figs.
\ref{Fig4_x}-\ref{Fig52_x} still persist. However, already at
$\alpha/\beta \sim 10^{-2}$ all vortex solutions disappear and only
stripe phase solutions are stable. This is illustrated in Fig.
\ref{Fig7_x}, where we have set $\beta = 1~\mu$eV$\mu$m and
$\alpha/\beta = 1/2$ for a spatially uniform condensate as a initial
condition of the imaginary time method.

\section{Conclusions}
We studied the stationary solutions describing various topological defects in the system of spinor indirect excitons applying the imaginary time method to the set of Gross-Pitaevskii equations. We analyzed the role of the SOI of Rashba and Dresselhaus types in formation of single vortices, half vortices and half vortex-antivortex pairs, and described the transition between warped vortex and stripe phase solutions in the presence of both Rashba and Dresselhaus SOI.

\emph{Acknowledgements.} We thank Yuri G. Rubo and Alexey V. Kavokin
for the valuable discussions. This work was supported by FP7 IRSES
projects SPINMET and POLAPHEN and Tier 1 project ``Polaritons for
novel device applixcations''. O. K. acknowledges the support from
Eimskip Fund. H. S. thanks Universidad Autonoma de Mexico for
hospitality.

\end{document}